\documentclass[amsfonts, amssymb, amsmath, reprint, showkeys, nofootinbib,superscriptaddress,  prx]{revtex4-2}

\usepackage{xcolor}
\usepackage{graphicx}
\usepackage[left=13mm,right=13mm,top=35mm,columnsep=15pt]{geometry} 
\usepackage{subfigure}
\usepackage{float}
\usepackage{adjustbox}
\usepackage{placeins}
\usepackage[T1]{fontenc}
\usepackage{csquotes}
\usepackage[vcentermath]{youngtab}

\usepackage[nointegrals]{wasysym}
\usepackage{amsmath}
\usepackage{amssymb}
\usepackage{amsthm}
\usepackage{braket}
\usepackage{bm}
\usepackage{mathtools}
\usepackage{tensor}
\usepackage{mathrsfs}

\def\openone{\leavevmode\hbox{\small$1$\normalsize\kern-.33em$1$}}

\newcommand{\abs}[1]{\left| #1 \right|} 
\newcommand{\avg}[1]{\left< #1 \right>}

\let\baraccent=\= 
\renewcommand{\=}[1]{\stackrel{#1}{=}}

\newcommand{\be}{\begin{equation}}
\newcommand{\bel}[1]{\begin{equation}\label{#1}}
\newcommand{\ee}{\end{equation}}

\newcommand{\ii}{\mathrm{i}}

\graphicspath{ {./images/} }
\graphicspath{ {./appendix_images/} }

\makeatletter

\renewenvironment{widetext@grid}{%
  \par\ignorespaces
  \setbox\widetext@top\vbox{%
   \vskip15\p@
   \hb@xt@\hsize{%
    \leaders\hrule\hfil
    \vrule\@height6\p@
   }%
   \vskip6\p@
  }%
  \setbox\widetext@bot\hb@xt@\hsize{%
    \vrule\@depth6\p@
    \leaders\hrule\hfil
  }%
  \onecolumngrid
  \let\set@footnotewidth\set@footnotewidth@ii
}{%
  \par
  \twocolumngrid\global\@ignoretrue
  \@endpetrue
}%

\makeatother
\usepackage[pdftex, pdftitle={Article}, pdfauthor={Author}]{hyperref}

\begin{document}
\title{Observation of a Prethermal \textit{U}(1) Discrete Time Crystal}
    \author{Andrew Stasiuk}
     \email{astasiuk@mit.edu}
    \affiliation{Department of Nuclear Science and Engineering, Massachusetts Institute of Technology, Cambridge, Massachusetts 02139, USA}

    \author{Paola Cappellaro}
    \affiliation{Department of Nuclear Science and Engineering, Massachusetts Institute of Technology, Cambridge, Massachusetts 02139, USA}
    
\date{\today} 

\begin{abstract}
A time crystal is a state of periodically driven matter which breaks discrete time translation symmetry. Time crystals have been demonstrated experimentally in various programmable quantum simulators and exemplify how non-equilibrium, driven quantum systems can exhibit intriguing and robust properties absent in systems at equilibrium. These robust driven states need to be stabilized by some mechanism, with the preeminent candidates being many-body localization and prethermalization. This introduces additional constraints that make it challenging  to experimentally observe time crystallinity in naturally occurring systems.    
Recent theoretical work has developed the notion of prethermalization \textit{without temperature}, expanding the class of time crystal systems  to explain time crystalline observations at (or near) infinite temperature. In this work, we conclusively observe the emergence of a prethermal $U(1)$ time crystalline state at quasi-infinite temperature in a solid-state NMR quantum emulator by verifying the requisites of prethermalization without temperature. In addition to observing the signature period-doubling behavior, we show the existence of a long-lived prethermal regime whose lifetime is significantly enhanced by strengthening an emergent $U(1)$ conservation law. Not only do we measure this enhancement through the global magnetization, but we also exploit on-site disorder to measure local observables, ruling out the possibility of many-body localization and confirming the emergence of long-range correlations.

\end{abstract}

\maketitle

\section{Introduction}

    The proposal for the existence of time crystals posited that there exist systems which might break discrete time translation symmetry~\cite{wilczek2012quantum}, in analogy to spatial symmetry breaking in conventional crystals. However, it was quickly shown that such a feat is impossible for quantum systems at equilibrium~\cite{bruno2013impossibility}. In out-of-equilibrium systems, however, there is no fundamental restriction forbidding the emergence of time crystalline order~\cite{else2017prethermal}. 

    Specifically, in periodically driven many-body quantum systems, it has been demonstrated that there exist long-lived states of matter which break discrete time translation symmetry~\cite{else2017prethermal}. To engineer such robust driven states, there must be a mechanism to stop Floquet heating and avoid driving the system to infinite temperature. For a closed quantum system with disorder, many-body localization (MBL) is sufficient to prevent Floquet heating~\cite{zhang2017observation}. Thus, early experiments   sought to use this phenomenon, though doubts have been cast on initial attempts at observing MBL discrete time crystals (DTCs) in an ion trap quantum simulator due to the system's dimensionality. Recently, there have been more conclusive observations of discrete time crystals stabilized by MBL in superconducting qubit processors~\cite{zhang2017observation}. However, the phenomenon of MBL is hard to achieve experimentally as it requires both  strong disorder and short-range interactions. Such stringent conditions make the MBL DTC difficult to realize experimentally~\cite{lazarides2017fate}.

    Instead of MBL, a different mechanism to avoid Floquet heating can be pursued, namely, the phenomenon of prethermalization~\cite{machado2020long}. Notably, prethermal quantum systems exhibit lifetimes which are exponential in the Floquet driving frequency, without the need for any source of disorder~\cite{berges2004prethermalization, mori2018thermalization}. However, prethermal time crystals do require long-range interactions and low temperatures to meaningfully induce spontaneous symmetry breaking (SSB)~\cite{else2017prethermal}. Prethermal discrete time crystals (PDTC) have also been observed experimentally in quantum systems with high purity initial states due to low temperatures or hyperpolarization~\cite{kyprianidis2021observation,beatrez2023critical,frey2022realization}.

    As the explorations into out-of-equilibrium quantum systems have matured, many experimental platforms have demonstrated the emergence of time crystalline signatures under periodic driving. In addition to observations in small scale simulators such as ion traps~\cite{kyprianidis2021observation} and superconducting qubits~\cite{frey2022realization}, time crystalline order has been observed in large-scale devices such as dipolar NV ensembles~\cite{choi2017observation,beatrez2023critical} and solid-state NMR systems~\cite{rovny2018p,rovny2018observation} (See~\cite{else2020discrete} for a more complete list of recent theoretical and experimental progress across various systems).

    In the case of NMR and other high entropy simulators, there has been some confusion as to the cause of the observed emergent time crystalline state, as the system was too high temperature to exhibit the effects of spontaneous symmetry breaking, and not disordered enough to be stabilized by MBL~\cite{else2020discrete, rovny2018observation, rovny2018p}. However, recent work has shown that a third type of time crystal is possible, a prethermal $U(1)$ time crystal, allowing for prethermalization \textit{without} temperature~\cite{luitz2020prethermalization}. It is likely that some previous high temperature studies of time crystallinity have been instances of prethermal $U(1)$ DTCs. However, verification of this fact requires the measurement of local observables previously thought to be inaccessible in NMR and other large-scale systems restricted to global control~\cite{luitz2020prethermalization}. 

    In this work, we drive a macroscopic solid-state NMR system~\cite{peng2019prethermalization} to engineer time crystallinity, and verify that we observe a prethermal $U(1)$ time crystal by comparing local to global observables. Notably, we utilize a recently developed technique to leverage on-site disorder to measure spatially averaged truly local observables in  large quantum ensembles without universal or single site control~\cite{peng2022disorder,martin2022localtherm}.

\section{Methods}
\begin{figure}
        \centering
    \includegraphics[width=.5\textwidth]{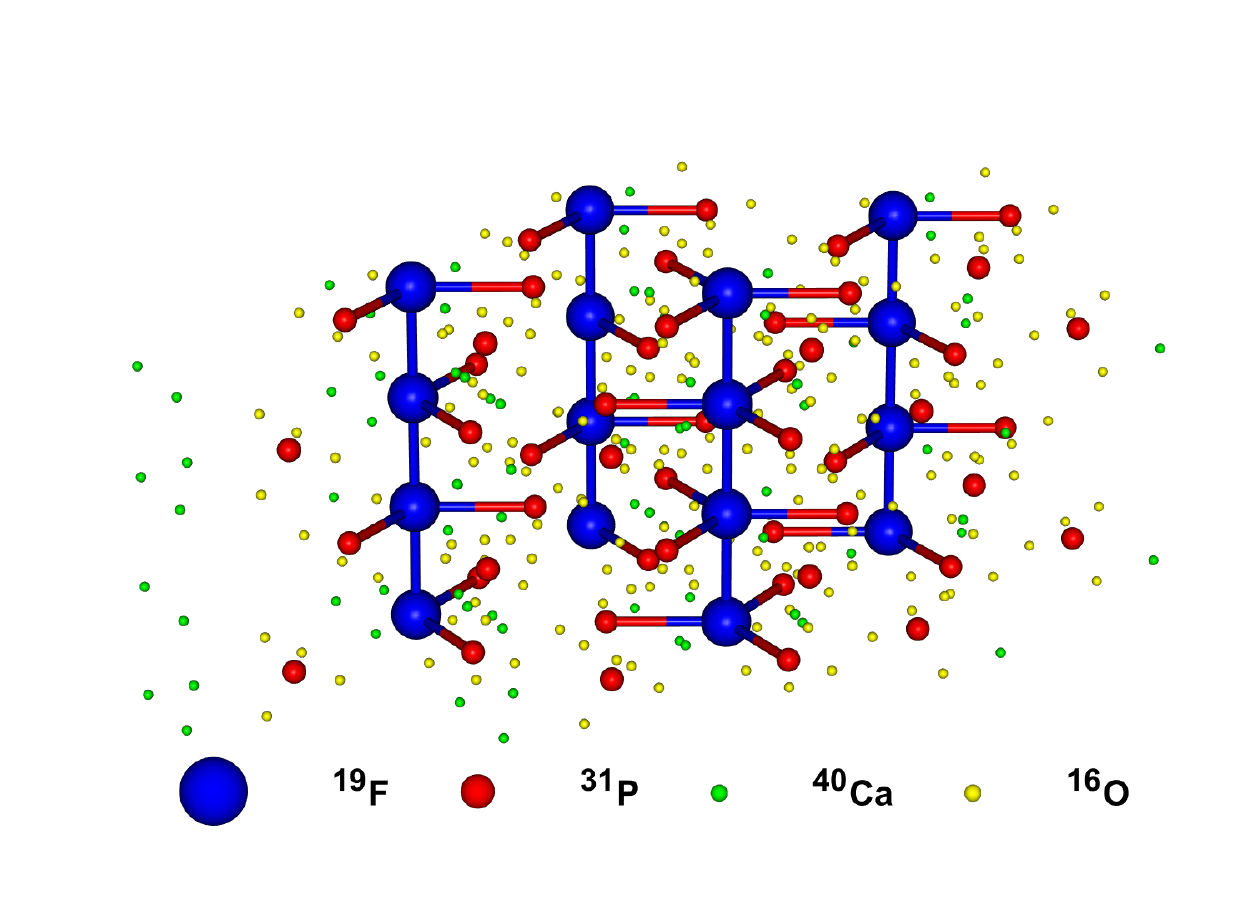}
        \caption{Crystal structure of fluorapatite, Ca$_5$(PO$_4$)$_3$F, with an emphasis on the spin-full nuclei. Intrachain fluorine (blue) bonds and nearest neighbor fluorine-phosphorus (blue-red) bonds are shown. Additionally, the spin-less nuclei of oxygen (yellow) and calcium (green) are included with no bonds shown. Spin-full isotopes of calcium and oxygen are less than 1\% natural abundance and neglected. The vertical crystal $c$-axis length is  6.805~\AA (with 2 $^{19}$F per unit cell), while the horizontal $a$,$b$-axes are  9.224~\AA\  long~\cite{leroy2001structure}.
        }
        \label{fig.crystal_structure}
    \end{figure}

    Our system is a quasi-1D nuclear spin ensemble. This is achieved experimentally using a single crystal sample of fluorapatite. The crystal structure of fluorapatite is given in figure \ref{fig.crystal_structure}.

    The crystal is placed in a large ($\sim$7.1 T) magnetic field aligned along the crystal's $c$-axis, so that the intrachain coupling is approximately 40 times larger than the interchain coupling~\cite{zhang2009nmr}. Within the rotating frame generated by the magnetic field, our system Hamiltonian has three main components: dipolar $z$, locally disordered $z$, and collective control.
    \begin{equation}
        \mathcal{H}(t) = \mathcal{H}_{Dz} + \mathcal{H}_{dis} + \mathcal{H}_c(t),
    \end{equation}
    Concretely, each component of the Hamiltonian is:
    \begin{align*}
        \mathcal{H}_{Dz} &= \frac{1}{2}\sum_{i<j} J_{ij}\big(\hat{S}_z^{(i)}\hat{S}_z^{(j)} - \frac{1}{2}(\hat{S}_x^{(i)}\hat{S}_x^{(j)}+\hat{S}_y^{(i)}\hat{S}_y^{(j)})\big)\\
        \mathcal{H}_{dis} &= \sum_{i} \omega_i \hat{S}_z^{(i)}\\
        \mathcal{H}_c(t) &= f(t) \big(\cos(\phi) \hat{S}_x + \sin(\phi)\hat{S}_y\big),
    \end{align*}
    where $J_{ij} = J_0/\abs{i-j}^3$, and we have defined the global magnetization operators
    \begin{equation}
        \hat{S}_{\nu} = \frac{1}{2}\sum_i \hat{\sigma}_{\nu}^{(i)} ,\,\,\, \nu\in\lbrace x, y, z\rbrace.
    \end{equation}
    The locally disordered field arises from the heteronuclear interaction of the fluorine and phosphorus spins. Finally, the collective control drive can be amplitude-modulated via the function $f(t)$ to produce square pulses of length $1.02 \mu$s resulting in $\pi/2$ rotations, with a minimum interpulse delay of $2.5 \mu$s. The transverse rotation axis is determined by the phase $\phi$, which has a resolution of 1 degree. Then, arbitrary x- and y-axis rotations are performed using the following decomposition,
    \begin{equation}
        R_y(\theta) = R_y(-\pi/2) R_z(\theta) R_y(\pi/2).
    \end{equation}
    z-axis rotations of an angle $\theta$ are performed by phase shifting all subsequent pulses by the same angle, $\theta$, called virtual Z-gate technique~\cite{knill2000algorithmic,mckay2017efficient}.

    As is standard in NMR, our initial state is well represented by the reduced density matrix $\delta\rho(0) = \hat{S}_z$. As the main observable of interest is also the magnetization along the $z$-axis, we can write the signal as an infinite temperature two-point correlator,
    \begin{equation}
        m_z(t) = \operatorname{tr}(\hat{\rho}(t)\hat{S}_z) \approx \operatorname{tr}(\delta\hat{\rho}(t)\hat{S}_z) = \avg{\hat{S}_z(t)\hat{S}_z}_{\beta=0}.
    \end{equation}
    By exploiting the dephasing induced by the disordered field, we can also generate states and observables that provide access to local correlations~\cite{peng2022disorder}. These disordered states are of the form
    \begin{equation}\label{eq.disorder_state}
        \delta\hat{\rho} = \sum_i \xi_i \hat{S}_z^{(i)},
    \end{equation}
    such that $\mathbb{E}[\xi_i \xi_j]\approx \delta_{ij}$.

   In order to demonstrate a $U(1)$ DTC we will engineer the simplest periodic sequence that can display such behavior. We will alternate between evolution under a Hamiltonian $\mathcal{H}_D$ (which might be obtained via Floquet Hamiltonian engineering, as described below) for a time $T$ and a collective rotation, $R_\nu(\theta)=e^{\ii \theta \hat{S}_{\nu}}$. We will often refer to the combination of these two steps as evolution with periodic kicking, in analogy to pushing a swing.
   
    \begin{figure}
        \centering
        \includegraphics[width=.48\textwidth]{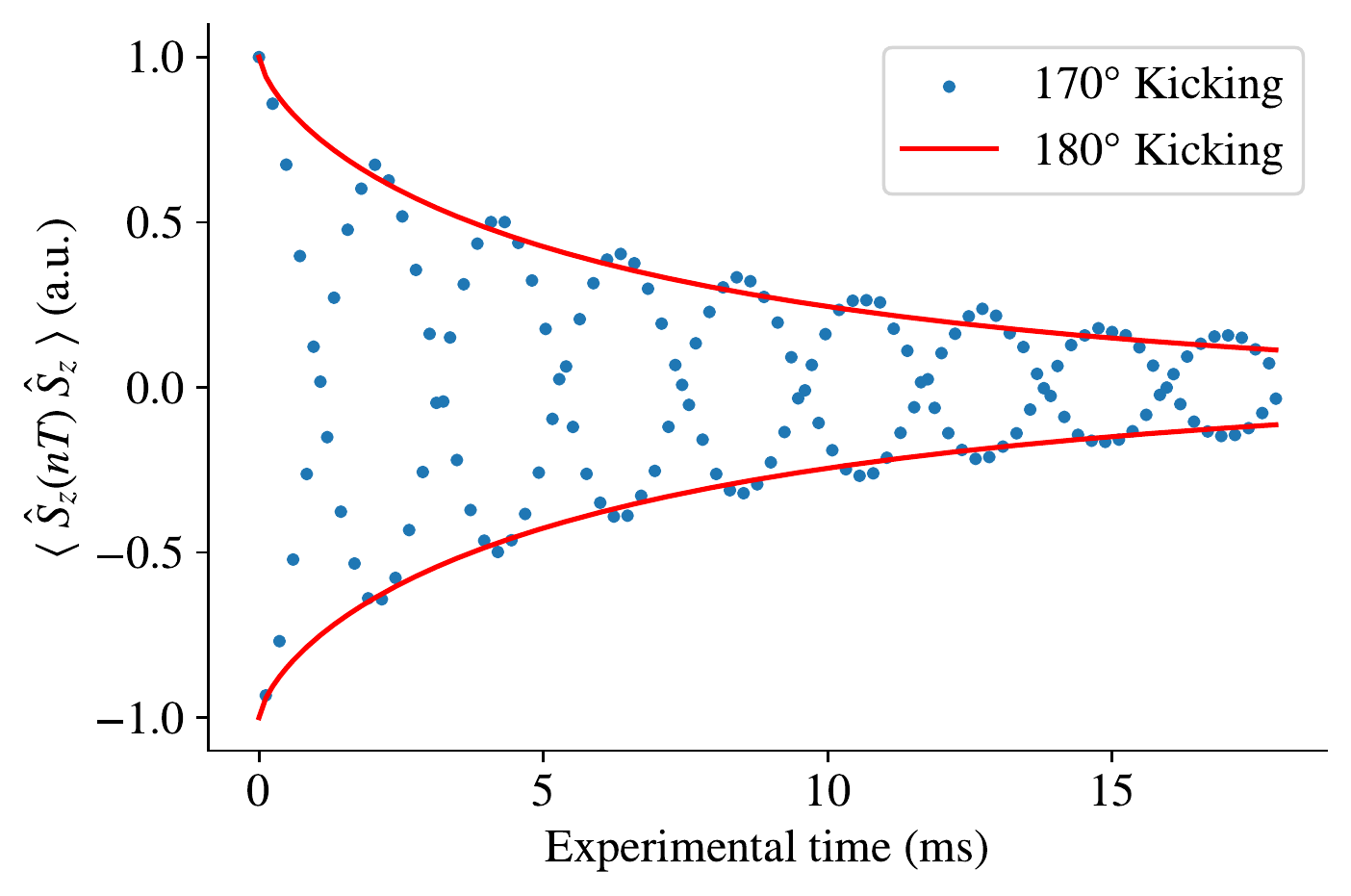} 
        \caption{Magnetization of non-interacting spin ensembles under repeated kicking. 
        Each Floquet cycle is $T=120\mu$s long, comprising evolution under the time-suspension sequence followed by a $R_y(\theta)$ unitary, with $\theta\in \lbrace 170, 180\rbrace$ degrees. Perfect $\pi$ kicking gives trivial period doubling, with a stretched exponential ($b=0.70)$ decay with timescale of  $\tau\approx6.573$~ms (Eq.~\ref{eq.non_int_decay}, $a\approx1$, $c\approx0$). The $\theta=(1-2\epsilon)\pi$ kicking results in beating with a frequency of $\frac{1}{2}\pm\epsilon$ decaying on a similar timescale.}
        \label{fig.non_interacting}
    \end{figure}

    We utilize the control to perform high-fidelity Floquet engineering sequences such as Wei16~\cite{wei2018exploring}, which allows rescaling the spin-spin interactions~\cite{Sanchez20}, and the Peng24 time-suspension sequence~\cite{peng2022deep}, which aims at canceling the whole Hamiltonian.  To illustrate this technique, and the fidelity of our control, we perform repeated $\theta=\pi$ and $(1-2\epsilon)\pi$ kicking on an effectively non-interacting ensemble of nuclear spins, $\mathcal{H}_D=0$. This is achieved using the Peng24 time-suspension sequence, which decouples interactions over a $T=120\mu$s Floquet period~\cite{peng2022deep}. As shown in figure \ref{fig.non_interacting}, the non-interacting decay envelope closely fits to an exponential with a sub-linear argument, given in equation \ref{eq.non_int_decay},    
    \begin{equation}\label{eq.non_int_decay}
        S(t) \equiv \avg{\hat{S}_z(t)\hat{S}_z} = a\exp(- (t/\tau)^b) + c.
    \end{equation}
    The long-lived signal ($\tau \approx 6573 \mu$s, compared to a free induction decay time of $\approx 30 \mu$s) and precise beating frequency confirms that the compiled pulse technique can perform an arbitrary high fidelity rotation in conjunction with a Hamiltonian engineering sequence.

\section{Results}

    \begin{figure*}
        \centering
        \includegraphics[width=.95\textwidth]{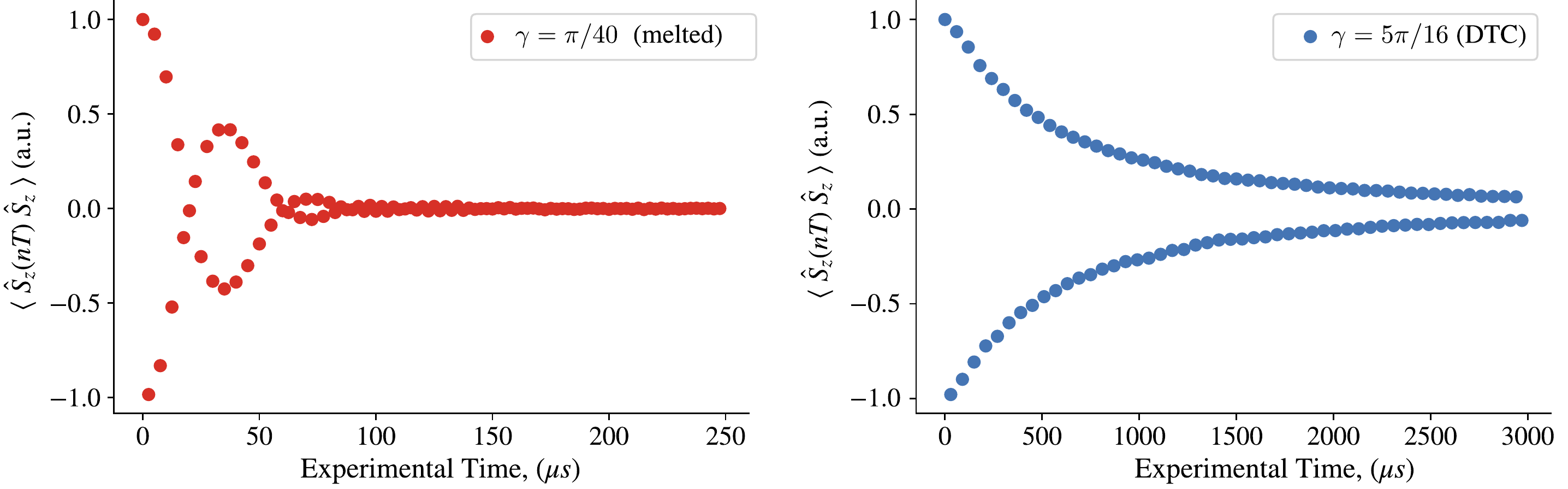}
        \caption{Time traces of the magnetization response for non-crystalline (left) and time crystalline (right) systems under $\theta=170^{\circ}$ driving. The interaction strength is set by the physical interaction time, $T=uT_0$. Notice the dramatic order of magnitude improvement in signal lifetime
        of the time crystal over the melted crystal.}
        \label{fig.time_trace}
    \end{figure*}
      
    We explore our system's ability to exhibit time crystallinity using a variety of techniques. First, we vary the effective Floquet period via two physical implementations to evaluate the robustness of our control scheme. Further, in an effort to distinguish between trivial  and non-trivial period doubling, we demonstrate our ability to generate robust period doubling as a function of the deviation from perfect $\pi$-kicking, $\epsilon$. To that end, we show the emergence of a long lived prethermal regime near the period doubling phase boundary. Finally, we conclusively diagnose our system as a prethermal $U(1)$ time crystal by showing its insensitivity to disordered fields and the rapid decay of local observables without beating. Given this diagnosis, we enhance $U(1)$ conservation by adding a stroboscopic $z$-field and demonstrate a significant lengthening of the prethermal timescale.

\subsection{Variable Interaction Time}\label{sec.var_time}
    Without loss of generality, to vary the effective kicking frequency $\omega$, we can either vary the interaction time between kicks or vary the interaction strength of $\mathcal{H}_D$ between equally spaced kicks. To this effect, we consider variations of the dimensionless \textit{interaction magnitude}, defined as
    \begin{equation}\label{eq.inter_mag}
        \gamma = u T_0 J_0,\,\,\, u \geq 0.
    \end{equation}
    In equation \ref{eq.inter_mag}, we take $T_0=120\mu$s and $J_0=32.843$ krad s$^{-1}$ to be fixed, while $u$ is variable  (here $T_0$ is chosen to  match the 120 $\mu$s Floquet period of Wei16 and Peng24.) Then, $J_0T_0 = 3.94$ radians, that is, $\gamma = 1.25 \pi u$.

    \begin{figure}[b]
        \centering
        \includegraphics[scale=0.65]{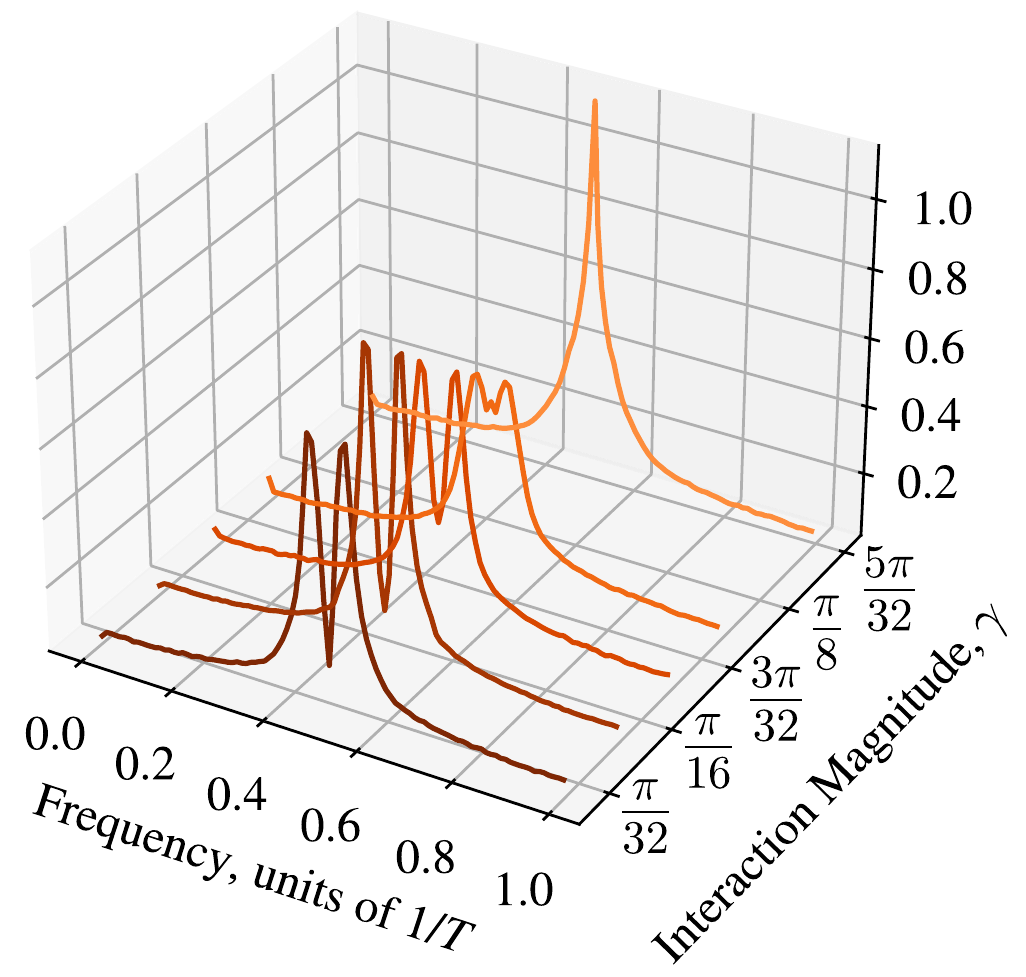}
        \caption{Fourier transform of the magnetization signal, $\mathcal{F}(m_z(t))$, for various interaction magnitudes $\gamma = u J_0 T_0$ under 170$^{\circ}$ kicking. The interaction magnitude is set by physically varying the time between consecutive $R_y(\theta)$ unitary kicks. For small $u$, the signals show beating consistent with the $\theta \neq \pi$ drive. At large enough $u$, the beating is replaced with a single ``period doubled'' peak, often used to herald a system's transformation to a time crystal.}
        \label{fig.per_3d_fourier}
    \end{figure}

    To explore the large $\gamma$ ($u>0.25$) regime, we particularize to $\theta=170^{\circ}$, and repeatedly drive the system with variable physical delay $T=uT_0$ between kicks. Hence, we vary $u$ by physically varying the interaction time, with $\mathcal{H}_D =\mathcal{H}_{Dz}$ corresponding to direct evolution of under the internal Hamiltonian. At large $u$, that is, low driving frequency, we find that the four-parameter model given in equation \ref{eq.non_int_decay} remains a good fit (fitting results for the magnetization are shown  in figure \ref{fig.verbose_fit_internal} of the Appendix.) The decay is close to a simple exponential, with the exponential parameter $b$  nearly unity, taking values between $0.92$ and $0.94$ for select values of $ u \in [0.25, 1.0]$. 
    We notice that the decay timescale as a function of $\gamma$, $\tau(\gamma)$, is linearly increasing (as expected as $\gamma$ becomes more dominant over $\pi-\theta$.) However, we find that the non-dimensional decay constant $\tau_F = \tau/T$ is bounded above by $\sim37$ Floquet periods (see fitting  details in the Appendix,  figure \ref{fig.verbose_fit_followup}.) This trend confirms that the increase in signal lifetime with increasing $\gamma$ is not a time crystalline effect, since time crystals have lifetimes which can be generically increased by stronger interactions. Thus, we determined that the period doubling signature for $u>.25$ is trivial, and not time crystalline.

    In the small $\gamma$ regime ($u \in (0, 0.25]$), we find a transition between beating and period-doubled response, and a corresponding breakdown in the fitting of the envelope to a single exponential. The real-time response of a time crystal and a melted crystal are shown in figure \ref{fig.time_trace}. In figure \ref{fig.per_3d_fourier}, we demonstrate our ability to induce period doubling in the magnetization signal by increasing $\gamma$, an oft-cited hallmark of time crystallinity. Further, in the small $\gamma$ regime where either beatings or period doubling occurs, we find that the single exponential model is a bad fit for the signal's decay. In fact, by transforming the observed signal $m_z(t)$ to $-\log{ \left|m_z(t)\right|}$, it becomes evident that there are two exponential time scales present, as shown in figure \ref{fig.verbose_fit_followup} of the appendix. We see a short and rapid decay followed by a much slower decay rate -- strong evidence for prethermalization and genuine time crystallinity. In fact, the slow decay rate of the prethermal state is expected to be exponentially suppressed by increasing driving frequency~\cite{luitz2020prethermalization,kyprianidis2021observation,else2017prethermal,zeng2017prethermal}. 
    
   In the current implementation, the driving frequency is experimentally limited.  Indeed, we note that $(1-2\epsilon)\pi$-pulse takes approximately 4.5 $\mu$s to perform. That is, the effective pulse length is longer than the interaction time of the smallest considered interaction strength of $u=.02$ ($T = uT_0 = 2.5 \mu$s). The interaction Hamiltonian during the drive can be a possible source of confounding error especially for small $u$. We eliminate this source of confounding error in the next section, by fixing $T$ and varying the effective interaction strength, $J = uJ_0$, directly.

    \begin{figure}[t]
         \centering
         \includegraphics[scale=0.6]{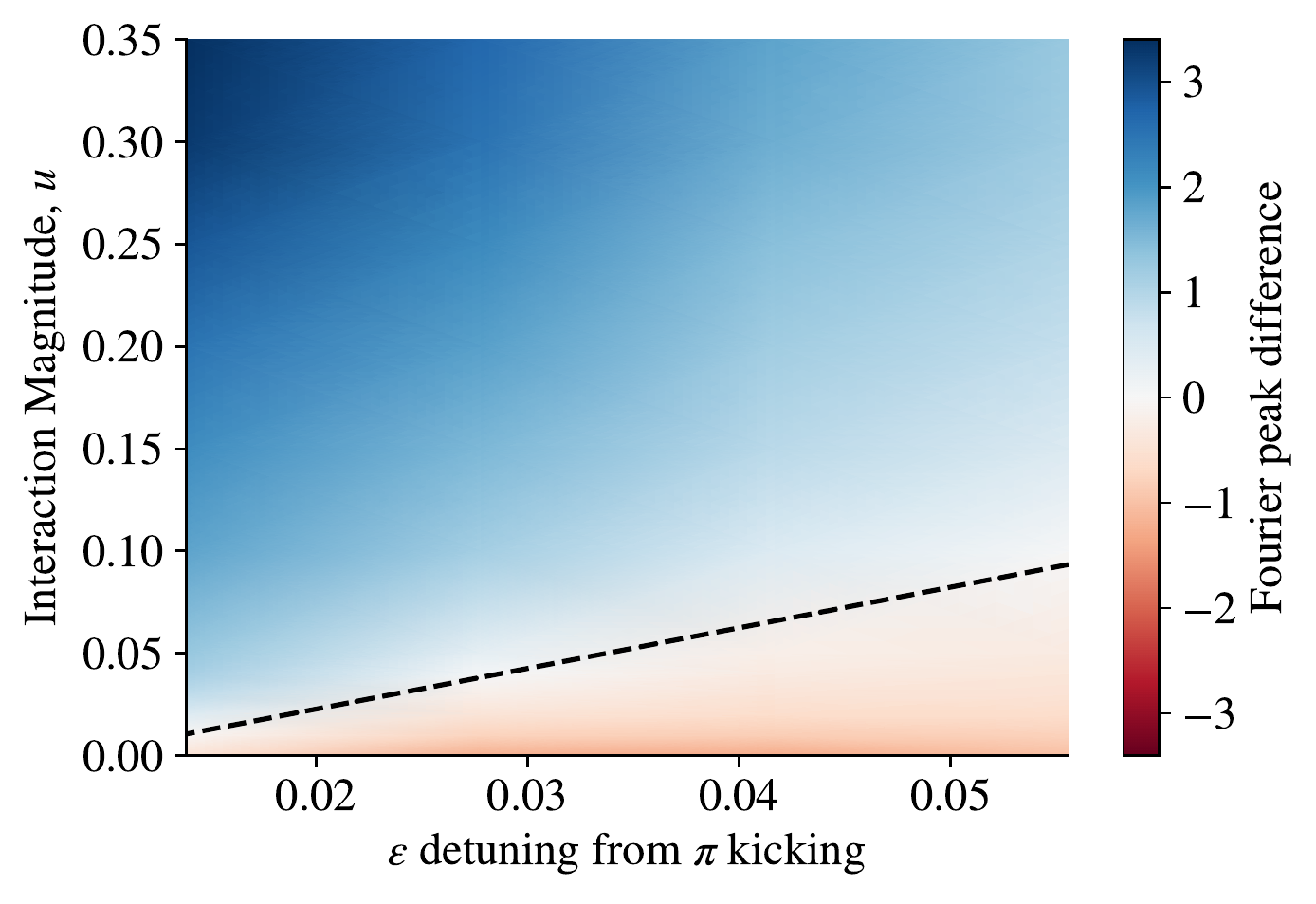}
         \caption{Difference in selected Fourier peak magnitudes of the observed magnetization under variable interaction magnitude ($u$) and kicking angle ($\epsilon$). Blue (red) shaded regions indicate a larger (smaller) period doubled response than beating. Using cubic spline interpolations, we estimate the intersection of beating and period doubling for each instance of $\epsilon$, corresponding to kicking angles $\theta\in\lbrace 160^{\circ}, 165^{\circ}, 170^{\circ}, 175^{\circ}\rbrace$. The resulting intersections lead to the fitted phase boundary, shown as a dashed black line.}
         \label{fig.phase_diagram}
     \end{figure}

    \subsection{Variable Interaction Strength}

     To reduce the effect of finite pulse width, we now utilize Hamiltonian engineering to fix the interaction time $T=T_0$ and vary the interaction strength $J_{\textrm{eff}} = uJ_0$. This allows for higher resolution and control in the interaction magnitude $u$, especially for smaller values of $u$. The range of interaction magnitudes allowed by Wei16 is limited and cannot achieve large values of $u$, namely $u \geq .35$ is forbidden. The large $u$ ($\gamma$) regime is well analyzed in the previous section, and in this section we intend to particularize only to the small $u$ regime.

     By varying the interaction magnitude $\gamma$ and the kicking angle $\theta=(1-2\epsilon)\pi$, we can estimate the boundary between signals displaying time translation symmetry breaking (TTSB) and trivial beating. To do so, we take the discrete Fourier transform of the measured signal. Given our signal $S(nT)$, for $n=0,1,\cdots,N-1$, the Fourier signal is given by
     \begin{equation}
         P(k) = \frac{1}{\sqrt{N}}\sum_{n=0}^{N-1} S(nT) e^{-\ii \frac{2\pi k n}{N}},
     \end{equation}
     where $k\in\lbrace0,1,\cdots,N-1\rbrace$. We find the existence of two main structures in the Fourier signal (figure \ref{fig.per_3d_fourier}): a peak at $f=\frac{1}{2}$, $P_{1/2} = \abs{P(N/2)}$, and a pair of peaks at $f=\frac{1}{2}\pm\epsilon$, $P_\epsilon = \abs{P(N/2 \pm N\epsilon)}$. Then, to estimate the transition boundary, we consider the difference $P_{1/2}-P_\epsilon$. The values of this metric are plotted in figure \ref{fig.phase_diagram} for various values of $u$ and $\epsilon$.
     
     Using this metric, we experimentally determine a ``phase boundary'' between a TTSB and beating magnetization signal, as done previously~\cite{pizzi2021higher,giachetti2022fractal}. In light of recent analyses of time crystallinity in NMR systems, this metric is insufficient to conclusively characterize time crystallinity~\cite{else2020discrete}. However, we can conclude that below the fitted boundary line at $P_{1/2}-P_\epsilon=0$, beating is dominant, and thus the observed behavior is not time crystalline. Instead, in the blue regions above the boundary line, figure \ref{fig.phase_diagram} simply indicates that the observed signal \textit{might be} time crystalline.

     As discussed in Section \ref{sec.var_time}, an interaction magnitude which is too large results in a trivial exponential decay, and thus not genuinely time crystalline. Further, we have carefully verified that an interaction magnitude which is too small also fails to be time crystalline. We expect, then, that genuine long-lived time crystals should emerge just past the crystallization boundary. In this regime, we expect to see a prethermal lifetime which is exponential in the driving frequency~\cite{luitz2020prethermalization,else2020discrete,peng2019prethermalization}. Namely, once prethermalization has occurred, we expect to find a significant increase in the in any present decay timescale.
     
\begin{figure*}
        \centering
        \includegraphics[width=.95\textwidth]{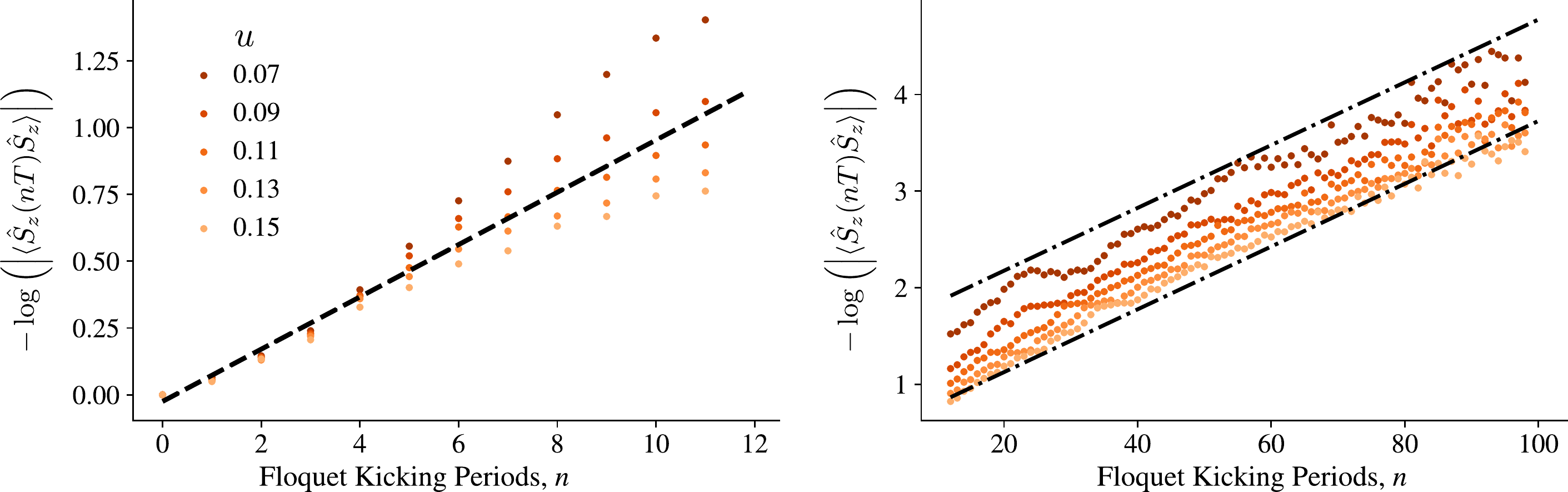}
        \caption{$-\log\left(\abs{\langle{\hat{S}_z(nT)\hat{S}_z}}\rangle\right)$ as a function of Floquet kicking periods, $n$, for $\theta=170^{\circ}$, and select interaction magnitudes $u$, in which period doubled behavior is dominant and the driving frequency is not too slow. Notably, we see the existence of two exponential timescales shown via the linear fits in black. A short timescale $\tau_1 \approx 7.2$ Floquet cycles (dashed line) present at early experimental times, shown on the left, gives way to a longer-lived timescale $\tau_2 \approx 30.8$ Floquet cycles (dot-dashed line) after an initial prethermalization period, shown on the right. The partition between early and late times is made to optimize the quality of the linear fits in both regimes.}
         \label{fig.two_timescale}
    \end{figure*}

     In figure \ref{fig.two_timescale}, we consider the magnetization signal for various interaction magnitudes in the regime where prethermalization is expected to occur. By fitting the observed signal to a two-timescale model, we are able to confirm the existence of a long lived prethermal state which occurs after a period of rapid prethermalization.

     Given the compelling evidence of prethermalization, robust period doubling, and extended lifetimes, we can conclude that these observations are genuinely time crystalline. It has been previously argued that at infinite temperature, and in short-range interacting 1D systems, spontaneous symmetry breaking (SSB) prethermal discrete time crystals are forbidden~\cite{else2017prethermal, else2020discrete, machado2020long}. In this section, we have demonstrated compelling evidence to support the observation of a prethermal $U(1)$ time crystal, which is robust and long lived even in a quasi-1D systems at an effective infinite temperature~\cite{luitz2020prethermalization}. In the following sections, we verify this indeed the case by enhancing the prethermal lifetime and conclusively ruling out MBL.

     \subsection{Behavior of Spatiotemporal Correlations}

    Further verification of time crystallinity and $U(1)$ prethermalization requires exploration of additional observables. In particular, the emergent $U(1)$ conservation law is \textit{global}. That is, our prethermal Hamiltonian conserves global, also called ``collective'', magnetization, but not local magnetization. This phenomenon has been previously understood in the context of spin diffusion in solid lattices, wherein initially local magnetization is transported through the lattice~\cite{boutis2004spin, zhang1998first}. In contrast, if the emergent time crystalline behavior was induced via MBL, then instead of global $U(1)$ conservation we should find an extensive number of local integrals of motion (LIOMs)~\cite{peng2019comparing,serbyn2013local}. Then, LIOMs will either conserve local magnetization or the local magnetization will oscillate between states of overlapping LIOMs~\cite{serbyn2014quantum}.

    The Wei16 engineering sequence, in addition to providing variable interaction magnitude in fixed time, allows us to modulate the strength of the disorder induced by nearby unpolarized phosphorus spins by varying the efficiency of refocusing the interaction. We exploited this capability to find no significant difference in the global magnetization's response to periodic driving from no disorder up to the maximum allowable disorder under Wei16, (see figure \ref{fig.disorder_investigation} in the appendix.) Thus, \textit{a priori} we expect our system to show no signs of many-body localization. However, there may be additional sources of disorder, such as drive field inhomogeneity outside of our primary model~\cite{rovny2018p}. To conclusively rule out MBL as the source of time-crystallinity, we utilize on-site disorder to produce a randomly polarized state (and observable), given by equation \ref{eq.disorder_state}. The procedure for creating such a state and observable were developed on this system and previously verified~\cite{peng2019prethermalization,peng2022disorder}. Notably, we measure truly local magnetization with an observable signal of the form,
    \begin{equation}
        S_{loc}(t) = \sum_i \avg{\hat{\sigma}_z^{(i)}(t)\hat{\sigma}_z^{(i)}}_{\beta=0}.
    \end{equation}

    \begin{figure}[t!]
        \centering
        \includegraphics[scale=0.6]{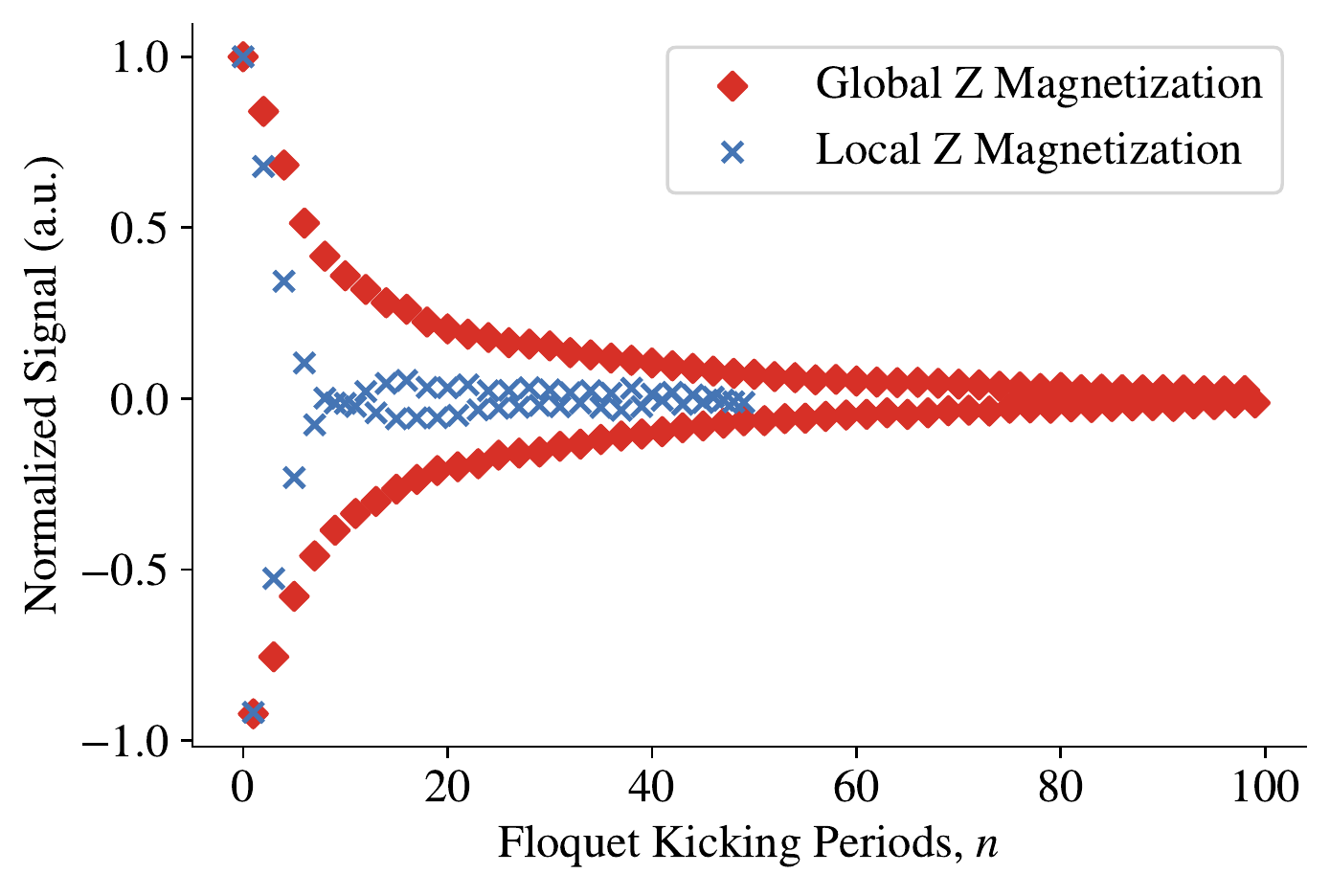}
        \caption{Decay of local and global magnetization signals for $\theta=170^{\circ}$ kicking with $\gamma=\pi/8$ ($u=0.1$.) The locally magnetized spin signal decays much more rapidly than the global one, as local magnetization is not conserved by the prethermal Hamiltonian. Further, there is no oscillatory behavior outside of period doubling, indicating no MBL eigenstates are induced.}
        \label{fig.local_v_global}
    \end{figure}

    In figure \ref{fig.local_v_global}, we illustrate the different behavior of local and global magnetization signals. Namely, we confirm that the local observable does not exhibit long-lived oscillations between local integrals of motion, and decays much more rapidly relative to the global signal. Thus, we confidently conclude that the observed time crystallinity is not induced by MBL. In the appendix, we show that this decay is due to coherent Hamiltonian evolution by demonstrating the long-time persistence of the state when interactions are turned off. Indeed, with no interactions the dynamics become trivial, single-spin dynamics, and we expect no difference between local and global correlations. Indeed, the signal persists for long times, with a similar decay profile to the one observed for the global magnetization in the non-interacting case (Fig.~\ref{fig.non_interacting}.) The details of this fitting are given in figure \ref{fig.nonint_local_comparison} of the appendix.

    The rapid decay of local magnetization further solidifies the notion that under time crystalline evolution, short-range correlated states evolve into long-range correlated states. This is evidenced by the fact that our global magnetization signal at long times must be dominated by non-local correlations, since $S_{loc}(t)\longrightarrow 0$ faster than $S(t)$ and thus the longer-lived signal,
    \begin{equation}
        S(t)-S_{loc}(t) = \sum_{i\neq j}\avg{\hat{\sigma}_z^{(i)}(t)\hat{\sigma}_z^{(j)}}_{\beta=0},
    \end{equation}
    is dominated by non-local correlations. 

    \subsection{Enhancing \textit{U}(1) Conservation}

    In an ideal closed system under the dipolar Hamiltonian $\hat{H}_{Dz}$, the magnetization in the $z$ direction is exactly conserved. Further, the anti-commutativity of $\exp(-\ii \pi \hat{S}_y)$ with $\hat{S}_z$ and commutativity of $\exp(-\ii \pi \hat{S}_y)$ with $\hat{H}_{Dz}$ leads to the long lived magnetization toggling response known as period doubling. This results in the emergent $U(1)$ conservation of the effective zeroth order two-period Floquet Hamiltonian. The emergent symmetry can be further enhanced by adding a field of strength $h$ in the $z$ direction, $\mathcal{H}_D=H_{Dz}+h\hat{S}_z$. Luitz \textit{et. al.} ~\cite{luitz2020prethermalization} have previously shown that the time-averaged effective Hamiltonian over two periods is given by,
    \begin{equation}\label{eq.effective_ham}
        \hat{H}_{\text{\textrm{eff}}} = \frac{T}{T+\epsilon}\bigg(\hat{H}_{Dz} + \frac{\epsilon_{\text{\textrm{eff}}}}{T}\hat{S}_y + h_{\text{\textrm{eff}}}\hat{S}_z\bigg) + O(1/\omega).
    \end{equation}
    Equivalent to considering the average two-period evolution, we could have chosen to consider a toggling frame in which we undo the $\pi$ rotation in each period, leaving only the  $\epsilon$ rotation. Corrections to the time-averaged Hamiltonian are of order $1/\omega \sim T$.  The parameters $h_{\text{\textrm{eff}}}$ and $\epsilon_{\text{\textrm{eff}}}$ are defined via the transformation into the toggling frame of the effective two-period Hamiltonian. So long as the effective magnitude of the $\hat{S}_y$ perturbation $\epsilon_\mathrm{eff}$ is small relative to the other energy scales of the Hamiltonian, namely $h_{\textrm{eff}} + J_0T_0$, the $\hat{S}_z$ magnetization is approximately conserved. The emergent $U(1)$ conservation is thus enhanced with decreasing $\epsilon_{\text{\textrm{eff}}}$ and increasing $h_{\text{\textrm{eff}}}$. Our phase diagram, figure \ref{fig.phase_diagram}, clearly demonstrates that the $U(1)$ conservation law is enhanced with decreasing $\epsilon$ at $h=0$. It remains to demonstrate the effect of non-zero $h$. To this end, we introduce a stroboscopic $\hat{S}_z$ field over the Floquet period $T$,
    \begin{equation}
        \mathcal{H}_D = h \hat{S}_z + u \hat{H}_{D_z},
    \end{equation}
    where the maximum field possible is restricted by periodicity, so that $0 \leq hT_0 \leq \pi$. 
    
    When the introduced $z$-field takes a maximal value of $hT_0=\pi$, we have that $h_{\text{\textrm{eff}}} = \epsilon_{\text{\textrm{eff}}} = 0$, even with $\epsilon\neq 0$.
    Intuitively, this result can be understood by focusing on the effects of the rotations on a single spin. During each period, the $z$-field results in a $R_z(\pi)$ rotation which inverts the axis of the applied $R_y(\theta)$ rotation.  Thus, to zeroth order, the $R_y$ rotations cancel out, and their effect will only enter to higher order. In contrast, all other choices of $0\leq hT_0<\pi$ leave a finite rotation around a transverse axis, resulting in magnetization which wraps around the Bloch sphere, only returning to its origin point when $n\theta = 0 \mod 2\pi$ (again neglecting higher-order terms in the Floquet expansion).

    We repeatedly perform the experimental and fitting procedures detailed in figure \ref{fig.two_timescale} for variable $hT_0=k\pi/4$ with $k~=~\{0,1,2,3,4\}$, for $\theta~=~170^{\circ}$. We record the two-timescale fitting results in table \ref{table.timescales} for each value in $hT_0$, making note of the $\gamma$ value at which the fitting was performed. Specific fitting details are given in the appendix in figure \ref{fig.two_timescale_verbose}. We note that the two-timescale signature of prethermalization is most pronounced near the period-doubling boundary, where the timescales differ by a factor of 5 or more. This trend is consistent with the fact that the prethermal lifetime increases with increasing driving frequency, $\omega \propto 1/\gamma$. Thus, the addition of the stroboscopic $z$ field increases the prethermal lifetime by stabilizing the time crystal at faster driving frequencies than would otherwise be achievable without the addition of a $z$ field.
    
    \begin{table}[thb]
    \begin{center}
        \begin{tabular}{| c | c || c | c | }
            \hline
            $hT_0$ & $\gamma$ 
            & $\tau_1$ (cycles) & $\tau_2$ (cycles)\\ [0.5 ex]
            \hline
            \hline
            $0$ & $\frac{7\pi}{80}$
            & 7.2 & 30.8 \\
            $\pi/4$ & $\frac\pi{16}$ 
            & 4.6 & 55.9 \\
            $\pi/2$ &$\frac\pi{16}$ 
            & 6.3 & 49.2 \\
            $3\pi/4$ &$\frac\pi{20}$ 
            & 8.7 & 49.8 \\
            $\pi$ &$\frac\pi{40}$ 
            & 10.6 & 77.5 \\ [1 ex]
            \hline
        \end{tabular}
    \end{center}
    \caption{Fitted timescales to prethermal $U(1)$ time crystalline magnetization signals under variable $z$-field and fixed $\theta = 170^{\circ}$.}
    \label{table.timescales}
    \end{table}

    Table \ref{table.timescales} indicates a strong enhancement of the timescale of the prethermal regime when a $z$-field is present. As previously developed, there are three notable regimes. The weakest conservation of $U(1)$ symmetry occurs for $hT_0=0$, corresponding to a prethermal timescale of about $31$ Floquet cycles. The introduction of a non-zero sub-maximal field, $0<hT_0<\pi$, causes the transverse field perturbation in the prethermal Hamiltonian to become increasingly off resonant. In these cases, the prethermal timescale is generically enhanced to around $50$ Floquet cycles. At maximal field, $hT_0=\pi$, the transverse field perturbation is averaged out in the toggling frame and the prethermal timescale is further enhanced to $77.5$ Floquet cycles. In real time units, this corresponds to a timescale of about 9.3 ms. The maximum field timescale is a dramatic improvement over the zero-field timescale, as expected~\cite{luitz2020prethermalization}.

\section{Conclusion}
    In this work, we have demonstrated the ability to create and melt time crystalline order in an infinite  effective temperature quasi-1D nuclear spin ensemble. 

    The DTC behavior leads to a characteristic \textit{period-doubling} response, which we analyzed by reconstructing an approximate phase diagram as a function of perturbation and driving rate. Noting however that the assumption that period doubling is a conclusive determinant for time crystallinity has been recently demonstrated to be false, we sought further proofs of time crystalline order. Indeed, we note that spontaneous symmetry breaking should be forbidden in our system. First, our experimental system can be considered quasi-one dimensional and so short-range, as the inter-chain and intra-chain couplings differ by a factor of about 40. Even if the assumption that our system is an ensemble of one-dimensional spin chains fails at longer times, and the dipolar interaction, $1/r^3$ can then be considered long-range, the high temperature still precludes time crystallinity due to prethermalization \textit{with temperature} necessary for a SSB PDTC. 

    We argue instead that our system fully demonstrates \textit{``prethermalization without temperature''}. 
    Notably, we utilized a new technique to verify the fast decay of local observables to conclusively rule out stabilization due to many-body localization, such that we observe a prethermal $U(1)$ time crystal. Thanks to this insight, we further introduced a $z$-field to successfully strengthen $U(1)$ conservation and significantly extend the time crystal's lifetime. 

    Our results also indicate that the time-crystalline phase is characterized by long-range correlations. An interesting future direction would be to try to utilize Hamiltonian engineering to perform time reversal on a time crystalline system, as done in previous NMR DTC experiments~\cite{rovny2018observation,rovny2018p}, and thus explore generated localization lengths. In a time crystalline system, we expect to measure rapidly growing localization lengths as the initial state evolves into a long-range spatiotemporal correlated state, in contrast to the local integrals of motion present in an MBL state.

\acknowledgments   
The authors would like to thank Pai Peng, Alex Ungar, and Santiago Hern\'andez-G\'omez for helpful discussions which enabled this work. This work was supported in part by the National Science Foundation under Grants No. PHY1915218.

\appendix
\section{Fitting to Period Doubled Decays, Variable Interaction Time}

\begin{figure*}
  \centering
    \includegraphics[width=0.44\textwidth]{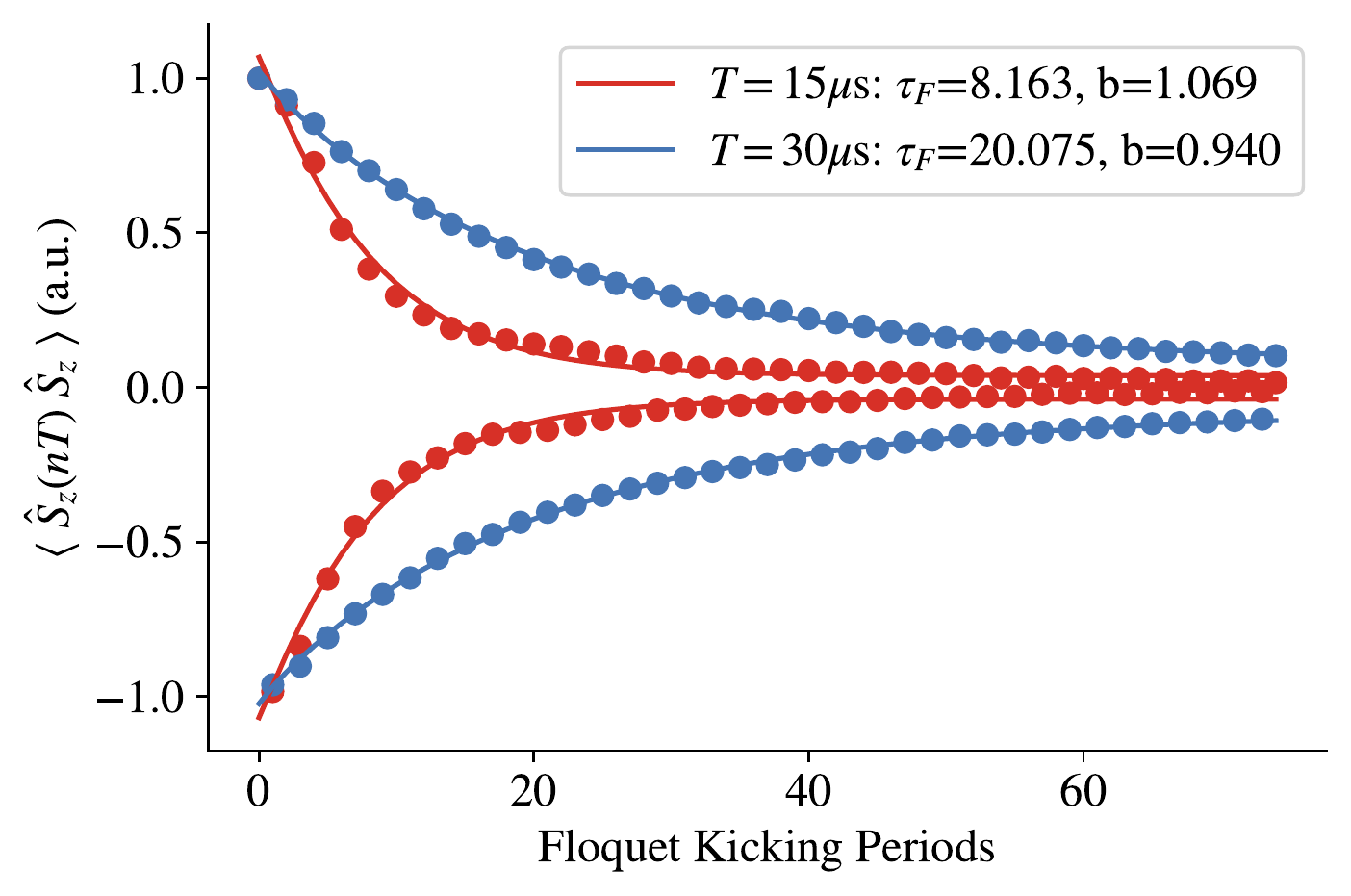}\qquad
      \includegraphics[width=0.44\textwidth]{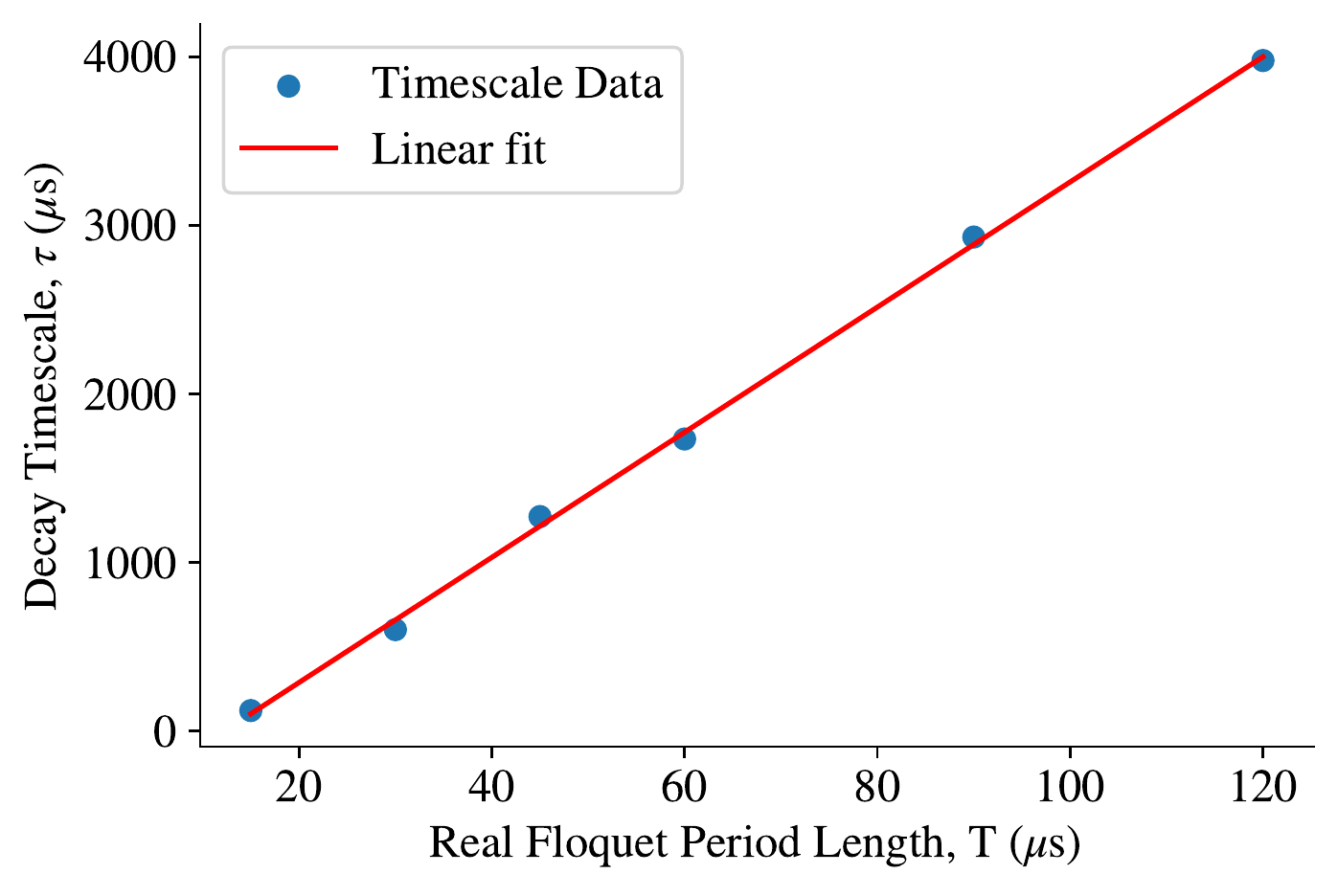}
  \caption{Left: Exemplary fitting results for variable interaction time DTC experiments. For each fit, the non-dimensional decay timescale $\tau_F$ (units of Floquet periods) is given along with the exponential stretching parameter $b$. (See SM for extended data). Right: Interpolation of the signal lifetimes extracted from left, as a function of the Floquet period $T$, $\tau(T) = aT - b$. The fitted parameters are $a = 37.11$ (non-dimensional) and $b=453.39\mu$s.}
  \label{fig.verbose_fit_internal}
\end{figure*}

Here, we detail the fitting procedure to the period doubled response under time crystalline driving, where the interaction period is variable. The model of interest is a four-parameter model,
\begin{equation}
        S(t) = a\exp(- (t/\tau)^b) + c,
\end{equation}
developed in equation \ref{eq.non_int_decay}. The fitting is performed in Python using SciPy's \verb+curve_fit+ function \cite{2020SciPy-NMeth}. We consider $T\in \lbrace 15, 30, 45, 60, 90, 120 \rbrace \mu$s, which corresponds to dimensionless interaction magnitudes $u\in\lbrace 0.125, 0.25, 0.375, 0.5, 0.75, 1.00\rbrace$. The results are shown in figure \ref{fig.verbose_fit_internal}.
\begin{figure}
  \centering
    \includegraphics[width=0.44\textwidth]{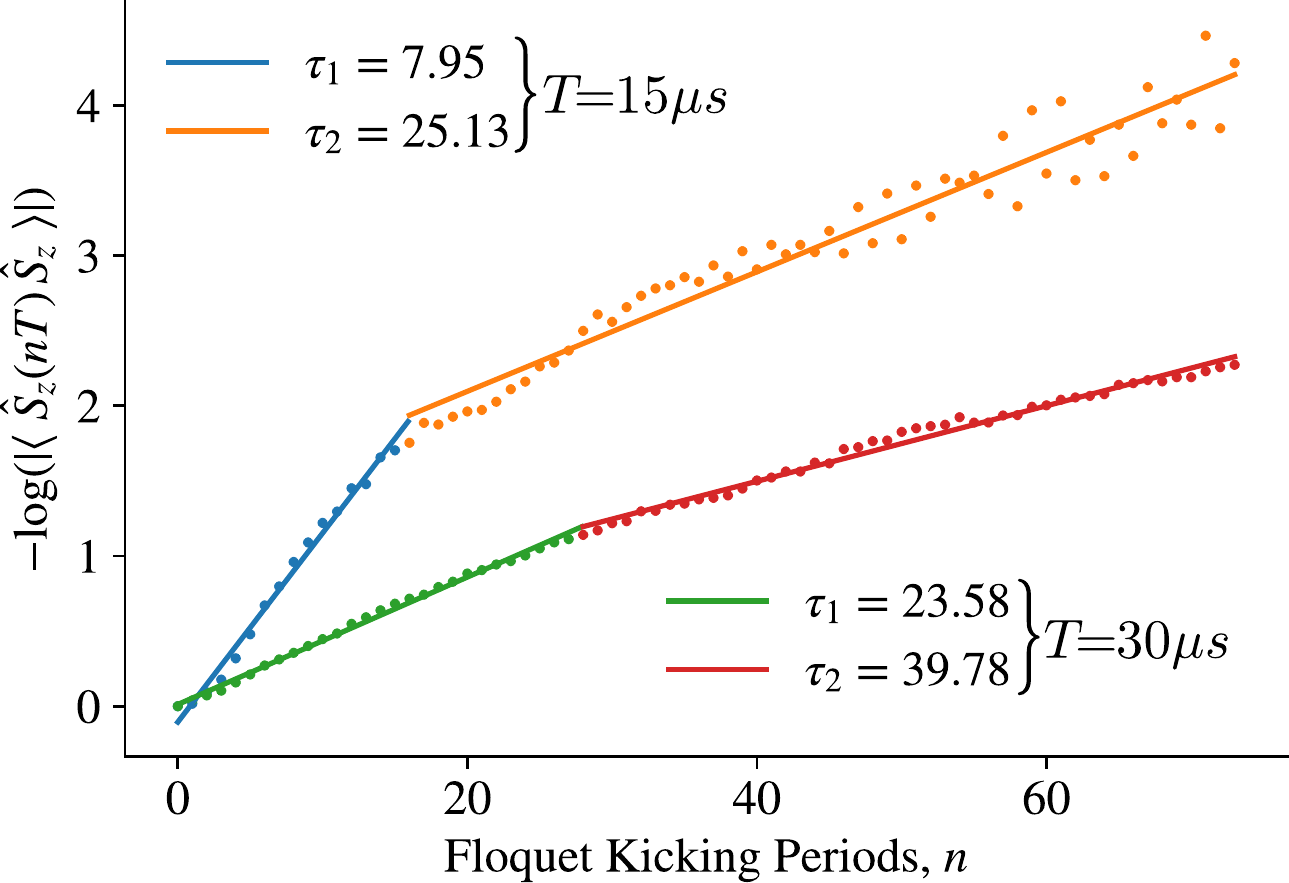}
  \caption{Two-timescale fitting for the $T=15\mu$s DTC experiment  and the $T=30\mu$s DTC experiment  in non-dimensional units of Floquet periods.}
  \label{fig.verbose_fit_followup}
\end{figure}

In figure \ref{fig.verbose_fit_internal}, there are two points of note. Firstly, the fitting for $T=15\mu$s shows systematic error -- the model is a bad fit for the given data. In this case, the two-timescale fitting procedure is a better choice and matches the notion that high driving frequency leads to prethermal time crystallinity. As $T$ increases, we similarly argue that the two-timescale fitting procedure is a bad choice. In figure \ref{fig.verbose_fit_followup}, we show the results of the two-timescale fitting procedure for $T=15,30\mu$s. The $T=15$ fitting demonstrates a strongly resolved transition between timescales, whereas the $T=30$ plot indicates that there might be a second timescale, but the transition between prethermalizing and prethermal timescales is weak. Notably, the choice of the cutting point between timescales does not significantly impact the fitting results. Hence, the argument in favor of prethermalization is quite weak, and so we take the single stretched exponential fit to be sufficiently explanatory. Secondly, there is a clear and apparent increase in the signal lifetime as the kicking period $T$ increases. In real-time units for the signal lifetime, this trend is linear, and is shown in figure \ref{fig.verbose_fit_internal} with a linear fitting.

Indeed, we see that the two-timescale fit well describes the $T=15\mu$s interaction time data, hence providing evidence for prethermalization, hence genuine time crystallinity. Further, the linear trend of the physical lifetimes, $\tau(T) = 37.11 T - 453.39$ ($\mu$s), implies that the non-dimensional Floquet period lifetime is  bounded. Indeed, $\tau_F(T) = 37.11 - 453.39/T$ (periods) is bounded above by 37.11 Floquet periods.

\section{Effects of Disordered Fields}

\begin{figure*}
  \centering
  \includegraphics[width=0.44 \textwidth]{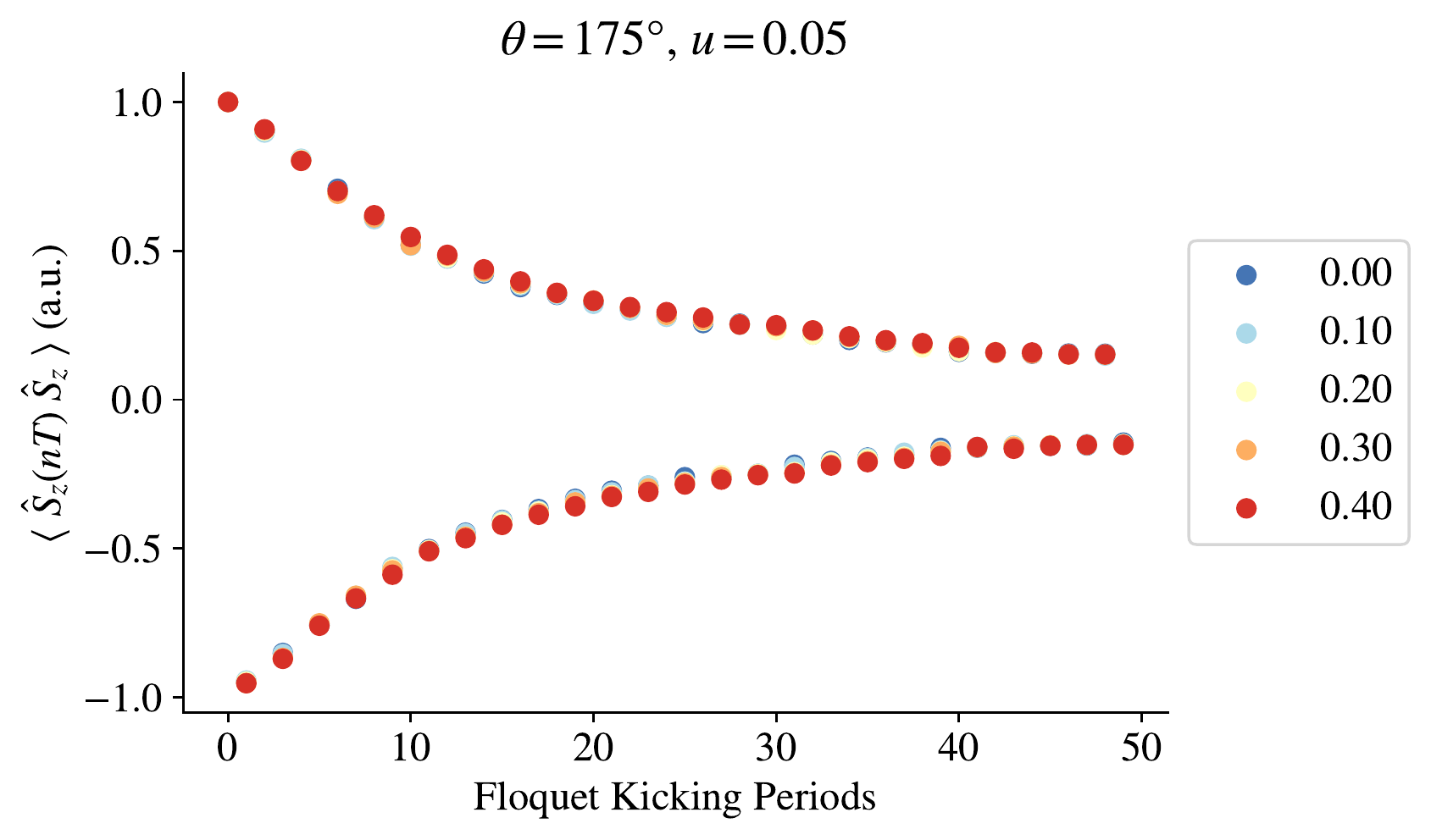}\qquad
  \includegraphics[width=0.44\textwidth]{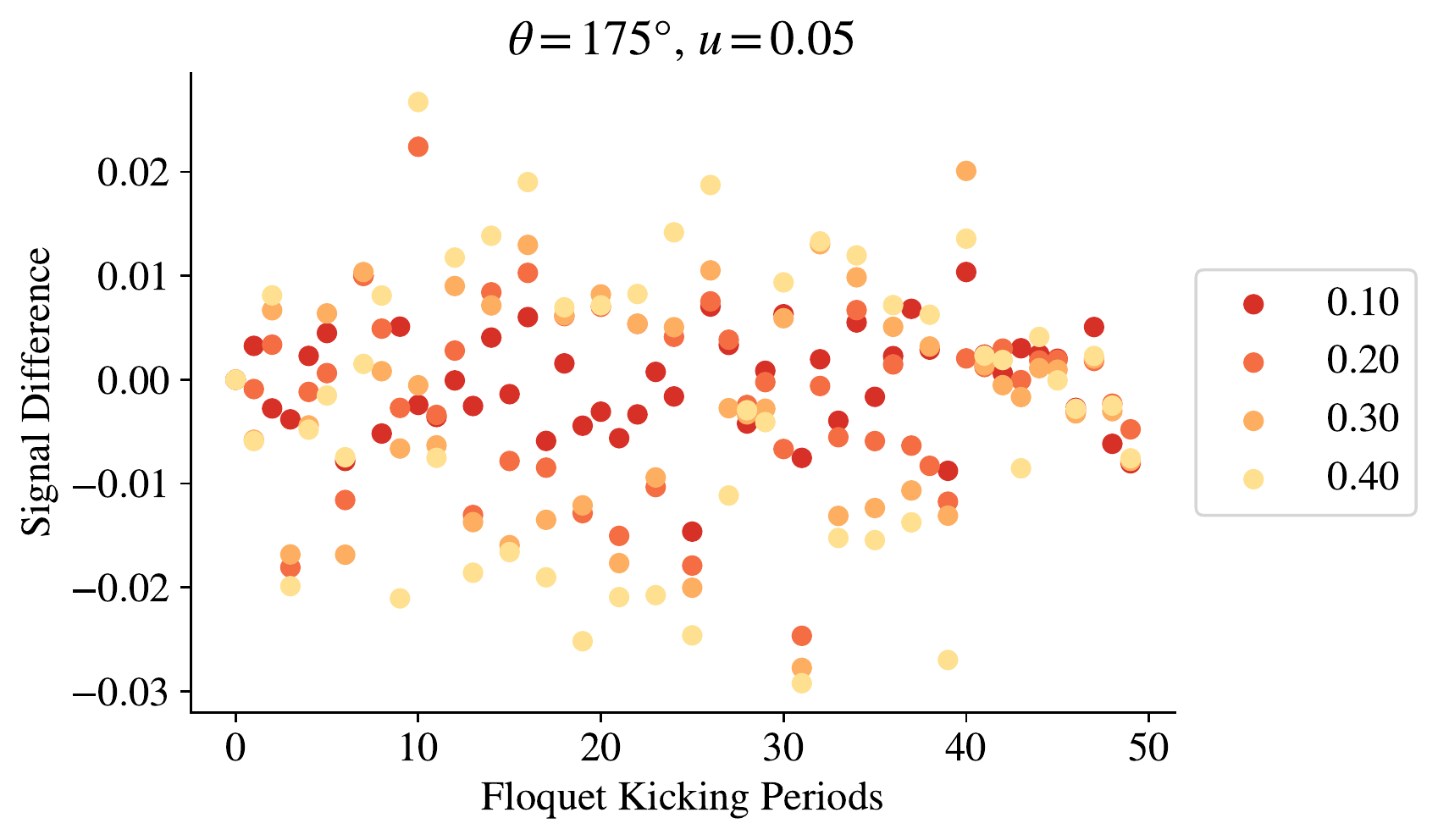}
   \includegraphics[width=0.44\textwidth]{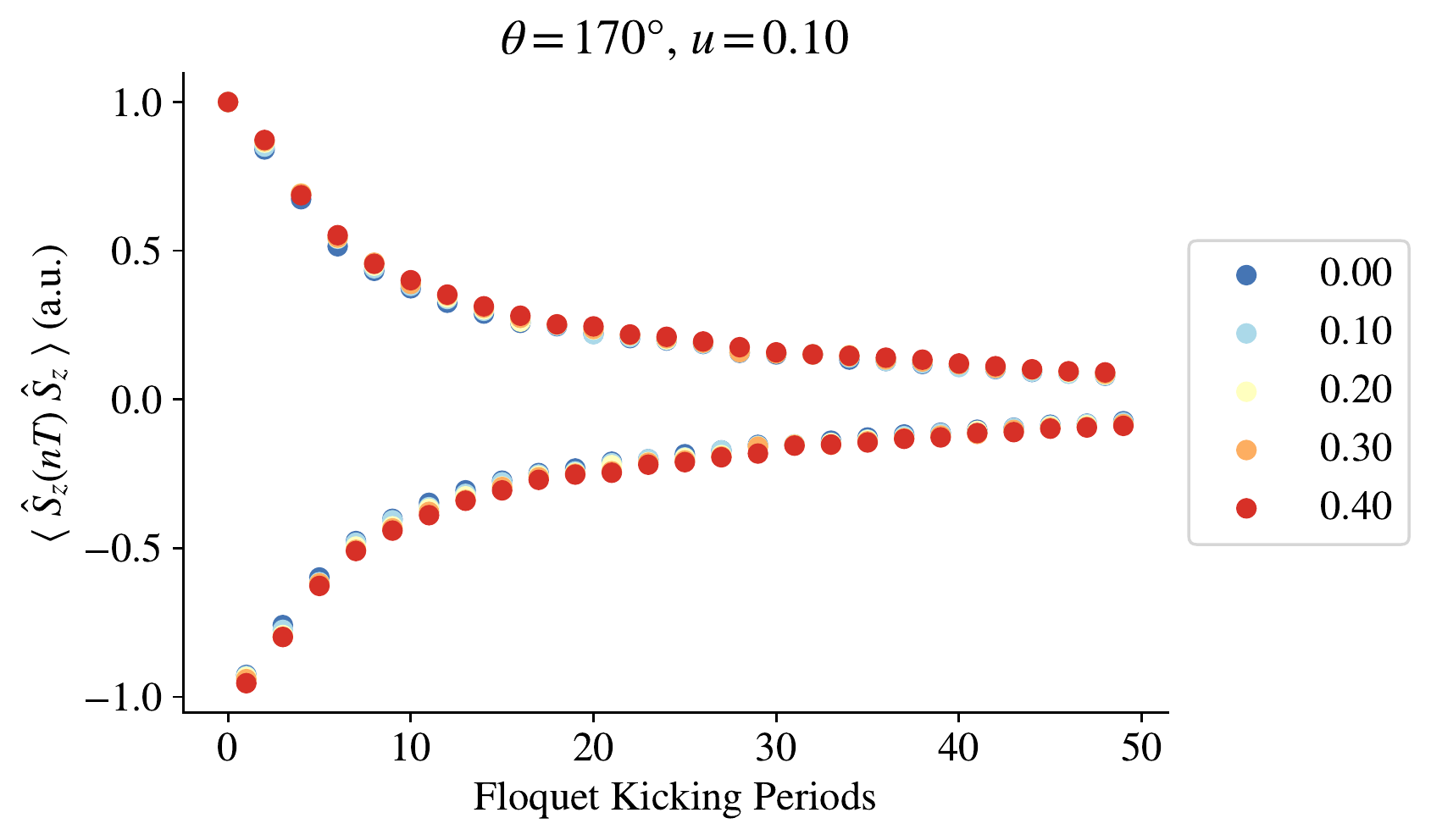}\qquad
    \includegraphics[width=0.44\textwidth]{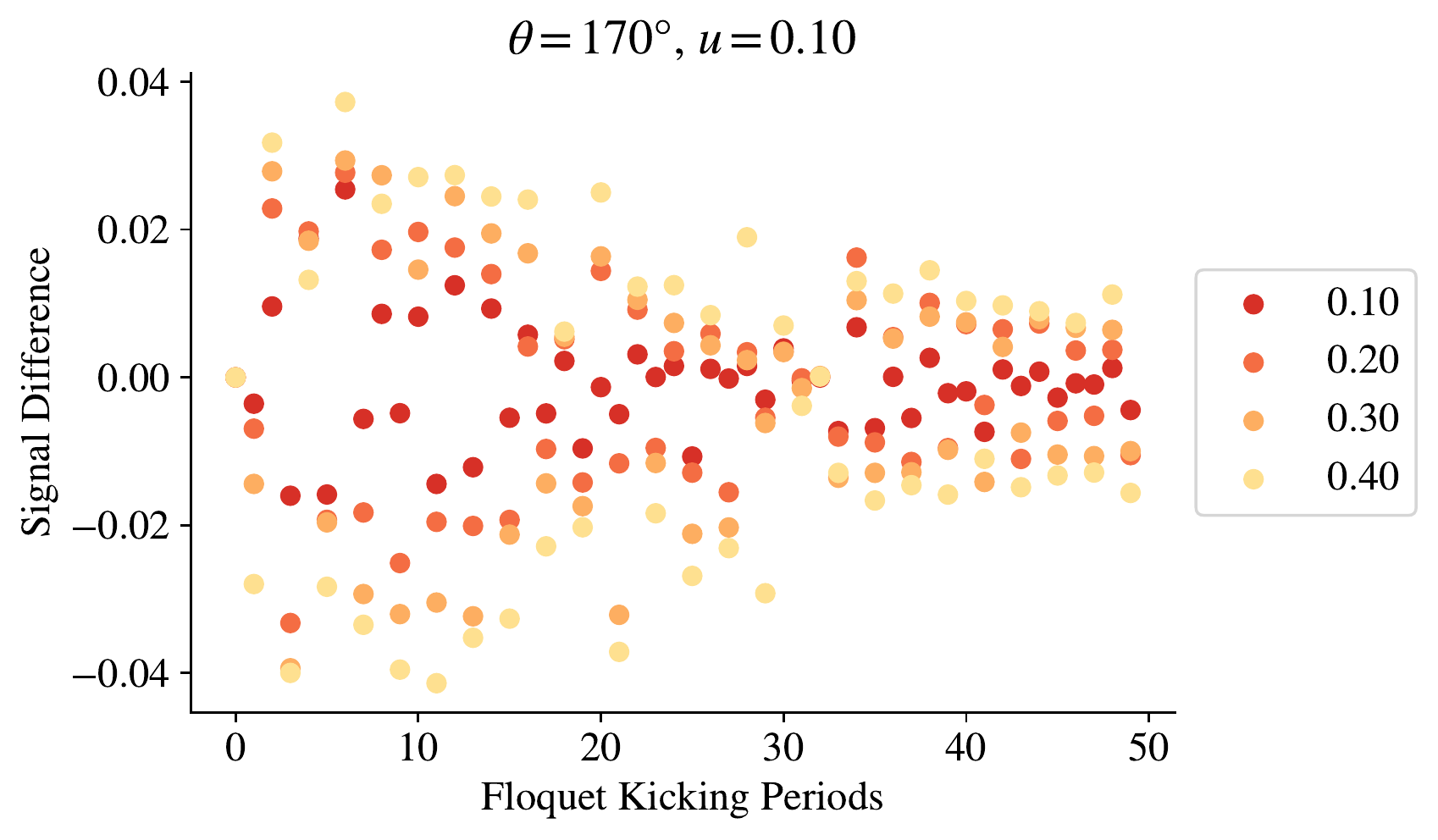}
    \caption{Effects of increasing disorder on time crystallinity for $\theta=175^{\circ}$, $u=.05$, and $\theta=170^{\circ}$, $u=0.08,0.10$. On the left, we show the measured magnetization signal for each disorder instance. On the right, we take the difference between experiments with disorder to the experiment with no disorder. Notice that the disordered field does not stabilize the signal; increasing disorder simply leads to increased variability in the measured signal.}
  \label{fig.disorder_investigation}
\end{figure*}

Using Wei16, we can vary the effective disorder strength over a single Floquet Period. Concretely, we have two tunable parameters, $u,v$, such that our Floquet Hamiltonian is 
\begin{equation}
    \hat{H}_{F} = \frac{v}{3} \sum_i r_i \hat{\sigma}_z^{(i)} +  \frac{u}{2} \sum_{i<j} J_0 \bigg(\hat{S}_z^{(i)}\hat{S}_z^{(j)} - \frac{1}{2}\big(\hat{S}_x^{(i)}\hat{S}_x^{(j)}+ \hat{S}_y^{(i)}\hat{S}_y^{(j)}\big)\bigg).
\end{equation}
Above, $r_i$ is the locally disordered field induced by the unpolarized phosphorus spins, and $0 \leq v \leq 0.4$ is the allowed parameter range under Wei16. Then, for selected kicking angles $\theta$, and interaction strengths $u$, we can investigate the effect of disorder on the stability of our time crystal. The results of this investigation are given in figure \ref{fig.disorder_investigation}. In summation, we see no evidence that increasing disorder leads to a longer-lived signal, hence MBL is likely not the stabilization mechanism for this time crystal.

\section{Local $z$ Magnetization State Manipulations}
\begin{figure*}
  \centering
    \includegraphics[width=0.4\textwidth]{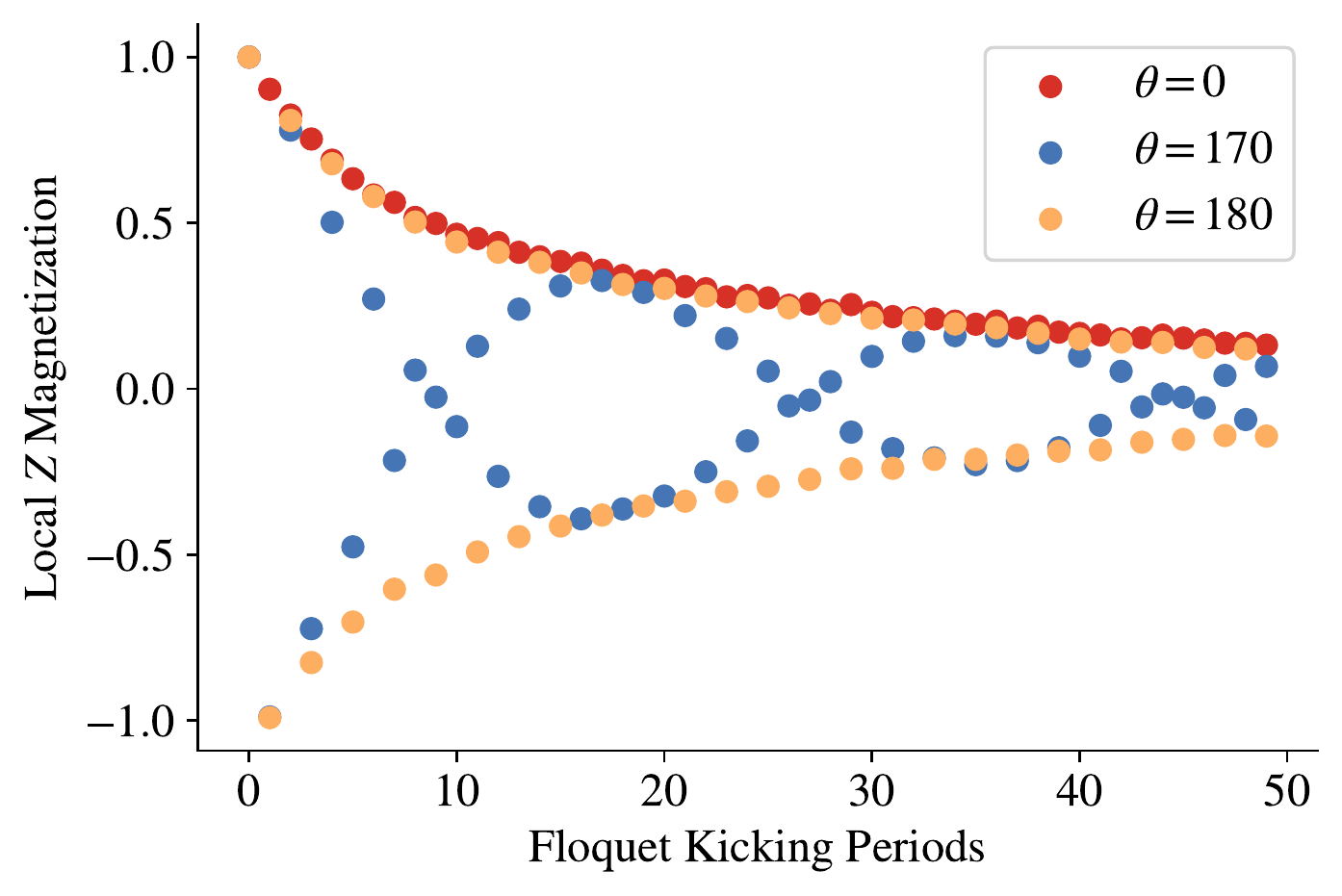}\qquad
    \includegraphics[width=0.4\textwidth]{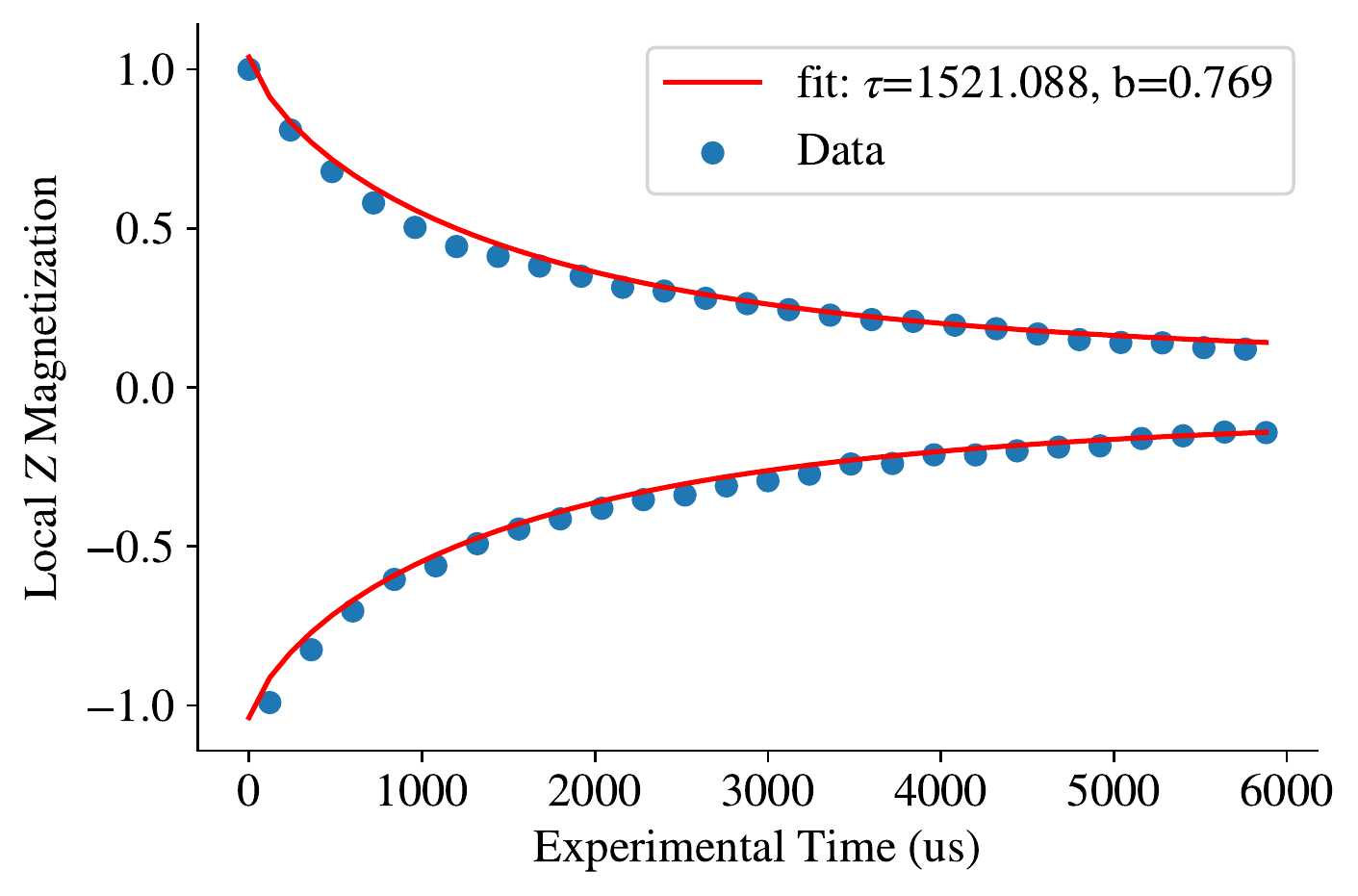}
  \caption{Results of local $z$ state experiments under periodic driving with interactions turned off via Peng24. On the left, signals are shown for kicking angles $\theta=0^{\circ},170^{\circ}, 180^{\circ}$. We notice that the decay envelope is similar for all three cases, as in the global $z$ case, albeit decaying more rapidly. On the right, we show the fitting of the $\theta=0^{\circ}$ signal to the four-parameter model given in equation \ref{eq.non_int_decay}.}
  \label{fig.nonint_local_comparison}
\end{figure*}

In order to further verify our ability to produce a robust local state, we consider the case where interactions are turned off entirely. Indeed, we want to ensure that the rapid decay observed in figure \ref{fig.local_v_global} is due to time crystal inducing interactions, and not due to poor state preparation. In figure \ref{fig.nonint_local_comparison}, we demonstrate that the local magnetization signal persists for long times when interactions are turned off with Peng24. In fact, the observed decay envelope fits well to the four-parameter model of \ref{eq.non_int_decay}, with a sub-linear exponential argument similar to the non-interacting global magnetization response. The time scale of the decay is much shorter for the local magnetization than global magnetization (1.5 ms vs 6.8 ms), but still quite long compared to $T_2 \approx 30\mu$s. In summation, the results of figure \ref{fig.nonint_local_comparison} demonstrate our state preparation procedure and control scheme are not the causes of the rapid decay seen in figure \ref{fig.local_v_global}.

\section{Two-timescale Fitting Results, Variable Interaction Strength}

\begin{figure*}
  \centering
  \includegraphics[width=0.4\textwidth]{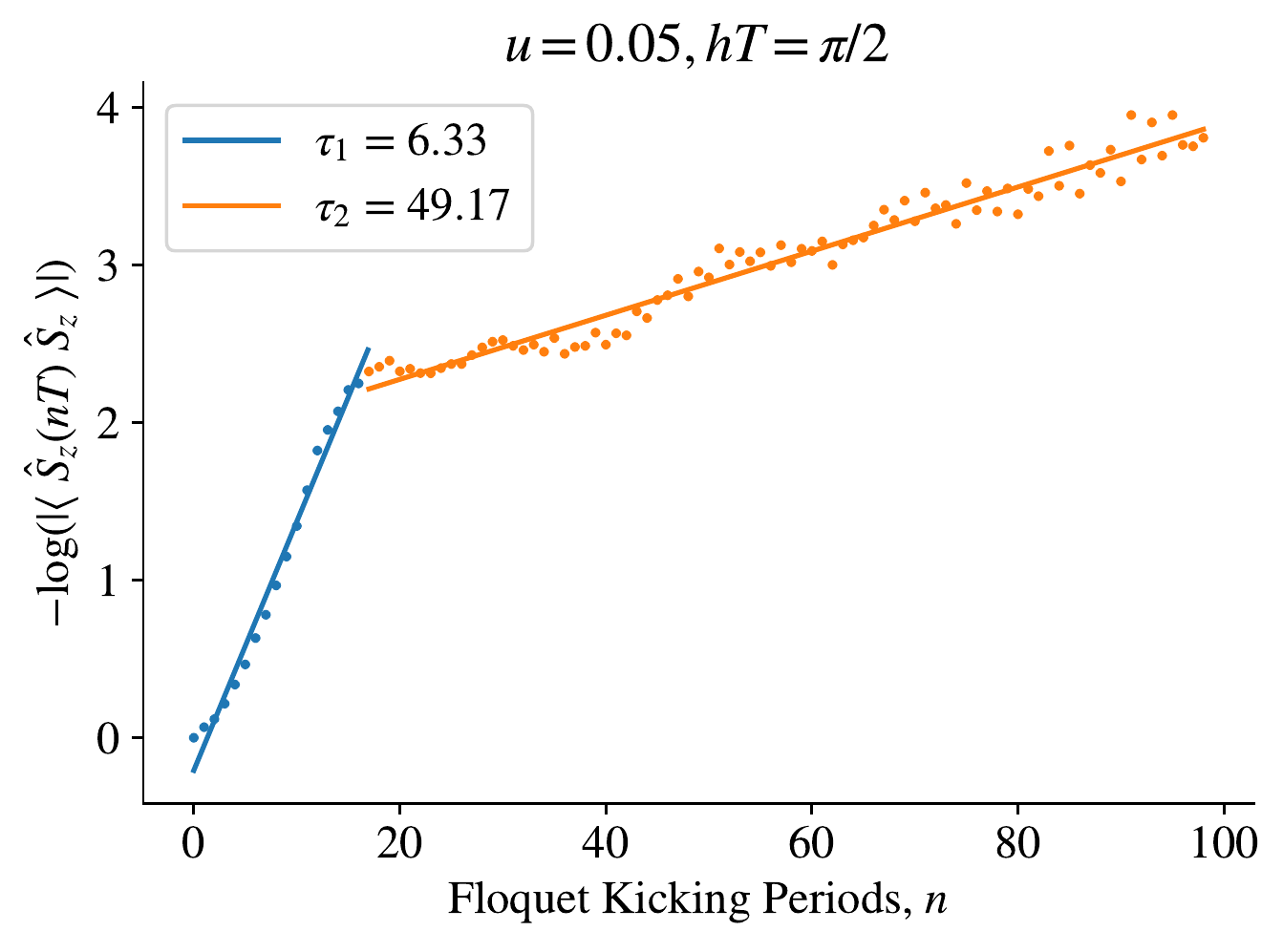}\qquad
  \includegraphics[width=0.4\textwidth]{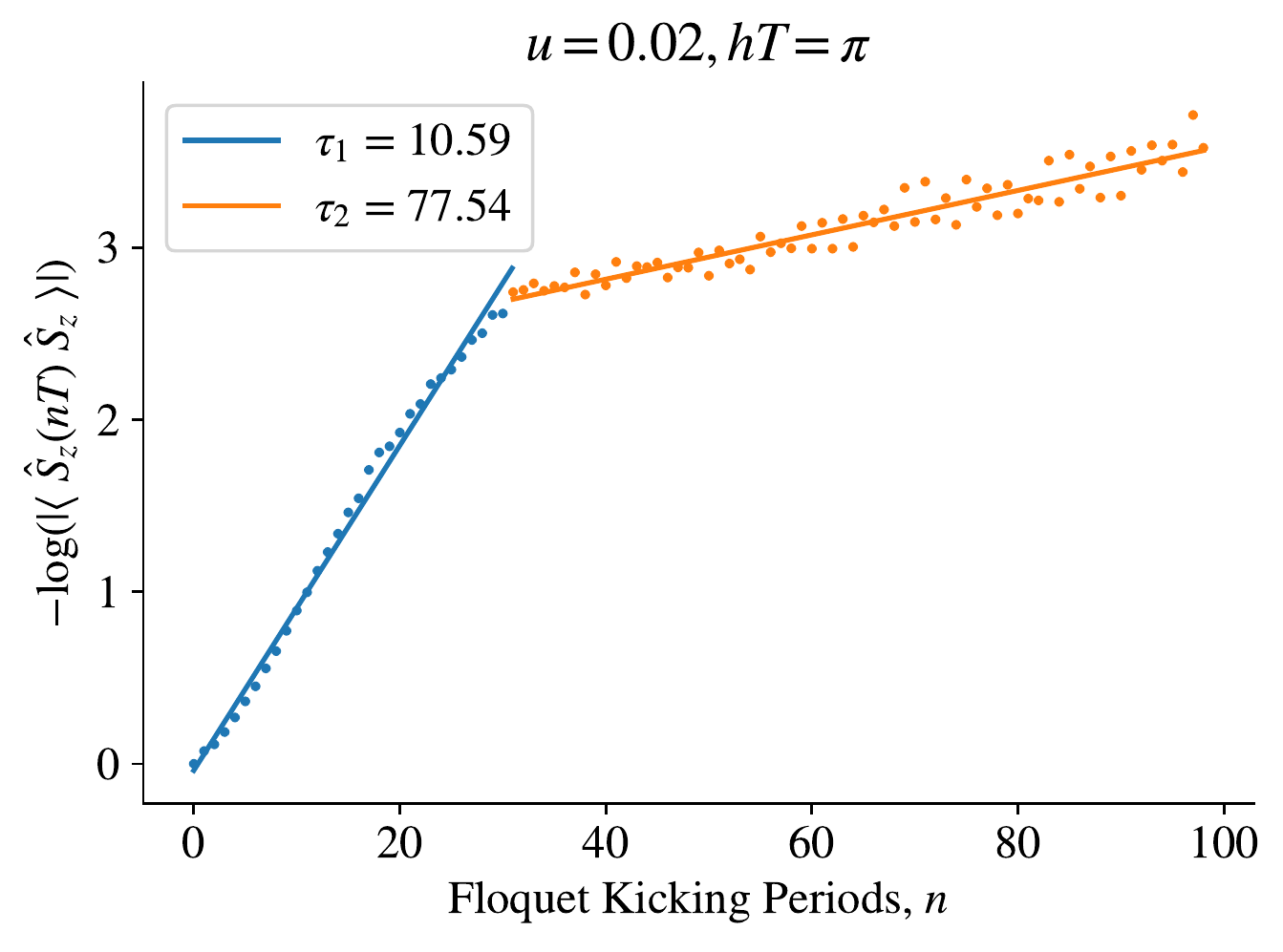}
\caption{Two-timescale fitting results for various applied stroboscopic $z$ fields, in non-dimensional units of Floquet periods. The data (see SM) generated by this procedure is used to populate the entries of table \ref{table.timescales}.} 
\label{fig.two_timescale_verbose}
\end{figure*}

In figure \ref{fig.two_timescale_verbose}, we show the two-timescale fitting results for stroboscopic $z$-field strengths of $hT = \pi/2, \pi$. This fitting procedure is performed for all $hT\in\lbrace \pi/4, \pi/2, 3\pi/4, \pi\rbrace$ to populate table \ref{table.timescales}. The fitting for $hT=0$ is shown in figure \ref{fig.two_timescale}. Concretely, $\tau_1$ gives the decay timescale of the initial state to the prethermal time crystalline state, and $\tau_2$ gives the lifetime of the prethermal time crystalline state. 

For each value of $hT$, an optimal $u$ is selected to maximize the two-timescale contrast. Generally, for larger values of $hT$, smaller values of $u$ are used. Given the theoretical expectation that $\tau_2$ depends exponentially on $u$, the quality of the fit is increasingly sensitive to large changes in $u$ with increasing $hT$. The fitting in figure \ref{fig.two_timescale} shows that the prethermal timescale is robust to changes in $u$, which can be attributed the fact that the transition to the time crystalline phase occurs near $u=0.07$. Thus, $\Delta u = 0.02$ is a small fraction of the interaction magnitude at the transition point. Even so, the contrast between timescales is most pronounced for $u=0.07$. However, for $hT=\pi$, the transition point to the time crystalline phase occurs near $u=0.02 (\gamma = \pi/40)$. Given our resolution in $u$ is limited to $0.01$, we are limited to (proportionally) large changes in the interaction magnitude relative to the transition point. If our timing control allowed for a higher resolution in $u$, then we could reconstruct a phase diagram for each $hT$ similar to figure \ref{fig.phase_diagram}. Qualitatively, we expect that the transition boundary be pushed down with increasing $hT$, thus increasing the phase space area occupied by the time crystalline state.

\newpage  
\bibliographystyle{apsrev}
\bibliography{refs}

\begin{thebibliography}{35}
\expandafter\ifx\csname natexlab\endcsname\relax\def\natexlab#1{#1}\fi
\expandafter\ifx\csname bibnamefont\endcsname\relax
  \def\bibnamefont#1{#1}\fi
\expandafter\ifx\csname bibfnamefont\endcsname\relax
  \def\bibfnamefont#1{#1}\fi
\expandafter\ifx\csname citenamefont\endcsname\relax
  \def\citenamefont#1{#1}\fi
\expandafter\ifx\csname url\endcsname\relax
  \def\url#1{\texttt{#1}}\fi
\expandafter\ifx\csname urlprefix\endcsname\relax\def\urlprefix{URL }\fi
\providecommand{\bibinfo}[2]{#2}
\providecommand{\eprint}[2][]{\url{#2}}

\bibitem[{\citenamefont{Wilczek}(2012)}]{wilczek2012quantum}
\bibinfo{author}{\bibfnamefont{F.}~\bibnamefont{Wilczek}},
  \bibinfo{journal}{Physical review letters} \textbf{\bibinfo{volume}{109}},
  \bibinfo{pages}{160401} (\bibinfo{year}{2012}).

\bibitem[{\citenamefont{Bruno}(2013)}]{bruno2013impossibility}
\bibinfo{author}{\bibfnamefont{P.}~\bibnamefont{Bruno}},
  \bibinfo{journal}{Physical review letters} \textbf{\bibinfo{volume}{111}},
  \bibinfo{pages}{070402} (\bibinfo{year}{2013}).

\bibitem[{\citenamefont{Else et~al.}(2017)\citenamefont{Else, Bauer, and
  Nayak}}]{else2017prethermal}
\bibinfo{author}{\bibfnamefont{D.~V.} \bibnamefont{Else}},
  \bibinfo{author}{\bibfnamefont{B.}~\bibnamefont{Bauer}}, \bibnamefont{and}
  \bibinfo{author}{\bibfnamefont{C.}~\bibnamefont{Nayak}},
  \bibinfo{journal}{Physical Review X} \textbf{\bibinfo{volume}{7}},
  \bibinfo{pages}{011026} (\bibinfo{year}{2017}).

\bibitem[{\citenamefont{Zhang et~al.}(2017)\citenamefont{Zhang, Hess,
  Kyprianidis, Becker, Lee, Smith, Pagano, Potirniche, Potter, Vishwanath
  et~al.}}]{zhang2017observation}
\bibinfo{author}{\bibfnamefont{J.}~\bibnamefont{Zhang}},
  \bibinfo{author}{\bibfnamefont{P.~W.} \bibnamefont{Hess}},
  \bibinfo{author}{\bibfnamefont{A.}~\bibnamefont{Kyprianidis}},
  \bibinfo{author}{\bibfnamefont{P.}~\bibnamefont{Becker}},
  \bibinfo{author}{\bibfnamefont{A.}~\bibnamefont{Lee}},
  \bibinfo{author}{\bibfnamefont{J.}~\bibnamefont{Smith}},
  \bibinfo{author}{\bibfnamefont{G.}~\bibnamefont{Pagano}},
  \bibinfo{author}{\bibfnamefont{I.-D.} \bibnamefont{Potirniche}},
  \bibinfo{author}{\bibfnamefont{A.~C.} \bibnamefont{Potter}},
  \bibinfo{author}{\bibfnamefont{A.}~\bibnamefont{Vishwanath}},
  \bibnamefont{et~al.}, \bibinfo{journal}{Nature}
  \textbf{\bibinfo{volume}{543}}, \bibinfo{pages}{217} (\bibinfo{year}{2017}).

\bibitem[{\citenamefont{Lazarides and Moessner}(2017)}]{lazarides2017fate}
\bibinfo{author}{\bibfnamefont{A.}~\bibnamefont{Lazarides}} \bibnamefont{and}
  \bibinfo{author}{\bibfnamefont{R.}~\bibnamefont{Moessner}},
  \bibinfo{journal}{Physical Review B} \textbf{\bibinfo{volume}{95}},
  \bibinfo{pages}{195135} (\bibinfo{year}{2017}).

\bibitem[{\citenamefont{Machado et~al.}(2020)\citenamefont{Machado, Else,
  Kahanamoku-Meyer, Nayak, and Yao}}]{machado2020long}
\bibinfo{author}{\bibfnamefont{F.}~\bibnamefont{Machado}},
  \bibinfo{author}{\bibfnamefont{D.~V.} \bibnamefont{Else}},
  \bibinfo{author}{\bibfnamefont{G.~D.} \bibnamefont{Kahanamoku-Meyer}},
  \bibinfo{author}{\bibfnamefont{C.}~\bibnamefont{Nayak}}, \bibnamefont{and}
  \bibinfo{author}{\bibfnamefont{N.~Y.} \bibnamefont{Yao}},
  \bibinfo{journal}{Physical Review X} \textbf{\bibinfo{volume}{10}},
  \bibinfo{pages}{011043} (\bibinfo{year}{2020}).

\bibitem[{\citenamefont{Berges et~al.}(2004)\citenamefont{Berges, Bors{\'a}nyi,
  and Wetterich}}]{berges2004prethermalization}
\bibinfo{author}{\bibfnamefont{J.}~\bibnamefont{Berges}},
  \bibinfo{author}{\bibfnamefont{S.}~\bibnamefont{Bors{\'a}nyi}},
  \bibnamefont{and}
  \bibinfo{author}{\bibfnamefont{C.}~\bibnamefont{Wetterich}},
  \bibinfo{journal}{Physical review letters} \textbf{\bibinfo{volume}{93}},
  \bibinfo{pages}{142002} (\bibinfo{year}{2004}).

\bibitem[{\citenamefont{Mori et~al.}(2018)\citenamefont{Mori, Ikeda, Kaminishi,
  and Ueda}}]{mori2018thermalization}
\bibinfo{author}{\bibfnamefont{T.}~\bibnamefont{Mori}},
  \bibinfo{author}{\bibfnamefont{T.~N.} \bibnamefont{Ikeda}},
  \bibinfo{author}{\bibfnamefont{E.}~\bibnamefont{Kaminishi}},
  \bibnamefont{and} \bibinfo{author}{\bibfnamefont{M.}~\bibnamefont{Ueda}},
  \bibinfo{journal}{Journal of Physics B: Atomic, Molecular and Optical
  Physics} \textbf{\bibinfo{volume}{51}}, \bibinfo{pages}{112001}
  (\bibinfo{year}{2018}).

\bibitem[{\citenamefont{Kyprianidis et~al.}(2021)\citenamefont{Kyprianidis,
  Machado, Morong, Becker, Collins, Else, Feng, Hess, Nayak, Pagano
  et~al.}}]{kyprianidis2021observation}
\bibinfo{author}{\bibfnamefont{A.}~\bibnamefont{Kyprianidis}},
  \bibinfo{author}{\bibfnamefont{F.}~\bibnamefont{Machado}},
  \bibinfo{author}{\bibfnamefont{W.}~\bibnamefont{Morong}},
  \bibinfo{author}{\bibfnamefont{P.}~\bibnamefont{Becker}},
  \bibinfo{author}{\bibfnamefont{K.~S.} \bibnamefont{Collins}},
  \bibinfo{author}{\bibfnamefont{D.~V.} \bibnamefont{Else}},
  \bibinfo{author}{\bibfnamefont{L.}~\bibnamefont{Feng}},
  \bibinfo{author}{\bibfnamefont{P.~W.} \bibnamefont{Hess}},
  \bibinfo{author}{\bibfnamefont{C.}~\bibnamefont{Nayak}},
  \bibinfo{author}{\bibfnamefont{G.}~\bibnamefont{Pagano}},
  \bibnamefont{et~al.}, \bibinfo{journal}{Science}
  \textbf{\bibinfo{volume}{372}}, \bibinfo{pages}{1192} (\bibinfo{year}{2021}).

\bibitem[{\citenamefont{Beatrez et~al.}(2023)\citenamefont{Beatrez,
  Fleckenstein, Pillai, de~Leon~Sanchez, Akkiraju, Diaz~Alcala, Conti,
  Reshetikhin, Druga, Bukov et~al.}}]{beatrez2023critical}
\bibinfo{author}{\bibfnamefont{W.}~\bibnamefont{Beatrez}},
  \bibinfo{author}{\bibfnamefont{C.}~\bibnamefont{Fleckenstein}},
  \bibinfo{author}{\bibfnamefont{A.}~\bibnamefont{Pillai}},
  \bibinfo{author}{\bibfnamefont{E.}~\bibnamefont{de~Leon~Sanchez}},
  \bibinfo{author}{\bibfnamefont{A.}~\bibnamefont{Akkiraju}},
  \bibinfo{author}{\bibfnamefont{J.}~\bibnamefont{Diaz~Alcala}},
  \bibinfo{author}{\bibfnamefont{S.}~\bibnamefont{Conti}},
  \bibinfo{author}{\bibfnamefont{P.}~\bibnamefont{Reshetikhin}},
  \bibinfo{author}{\bibfnamefont{E.}~\bibnamefont{Druga}},
  \bibinfo{author}{\bibfnamefont{M.}~\bibnamefont{Bukov}},
  \bibnamefont{et~al.}, \bibinfo{journal}{Nature Physics} pp.
  \bibinfo{pages}{1--7} (\bibinfo{year}{2023}).

\bibitem[{\citenamefont{Frey and Rachel}(2022)}]{frey2022realization}
\bibinfo{author}{\bibfnamefont{P.}~\bibnamefont{Frey}} \bibnamefont{and}
  \bibinfo{author}{\bibfnamefont{S.}~\bibnamefont{Rachel}},
  \bibinfo{journal}{Science advances} \textbf{\bibinfo{volume}{8}},
  \bibinfo{pages}{eabm7652} (\bibinfo{year}{2022}).

\bibitem[{\citenamefont{Choi et~al.}(2017)\citenamefont{Choi, Choi, Landig,
  Kucsko, Zhou, Isoya, Jelezko, Onoda, Sumiya, Khemani
  et~al.}}]{choi2017observation}
\bibinfo{author}{\bibfnamefont{S.}~\bibnamefont{Choi}},
  \bibinfo{author}{\bibfnamefont{J.}~\bibnamefont{Choi}},
  \bibinfo{author}{\bibfnamefont{R.}~\bibnamefont{Landig}},
  \bibinfo{author}{\bibfnamefont{G.}~\bibnamefont{Kucsko}},
  \bibinfo{author}{\bibfnamefont{H.}~\bibnamefont{Zhou}},
  \bibinfo{author}{\bibfnamefont{J.}~\bibnamefont{Isoya}},
  \bibinfo{author}{\bibfnamefont{F.}~\bibnamefont{Jelezko}},
  \bibinfo{author}{\bibfnamefont{S.}~\bibnamefont{Onoda}},
  \bibinfo{author}{\bibfnamefont{H.}~\bibnamefont{Sumiya}},
  \bibinfo{author}{\bibfnamefont{V.}~\bibnamefont{Khemani}},
  \bibnamefont{et~al.}, \bibinfo{journal}{Nature}
  \textbf{\bibinfo{volume}{543}}, \bibinfo{pages}{221} (\bibinfo{year}{2017}).

\bibitem[{\citenamefont{Rovny et~al.}(2018{\natexlab{a}})\citenamefont{Rovny,
  Blum, and Barrett}}]{rovny2018p}
\bibinfo{author}{\bibfnamefont{J.}~\bibnamefont{Rovny}},
  \bibinfo{author}{\bibfnamefont{R.~L.} \bibnamefont{Blum}}, \bibnamefont{and}
  \bibinfo{author}{\bibfnamefont{S.~E.} \bibnamefont{Barrett}},
  \bibinfo{journal}{Physical Review B} \textbf{\bibinfo{volume}{97}},
  \bibinfo{pages}{184301} (\bibinfo{year}{2018}{\natexlab{a}}).

\bibitem[{\citenamefont{Rovny et~al.}(2018{\natexlab{b}})\citenamefont{Rovny,
  Blum, and Barrett}}]{rovny2018observation}
\bibinfo{author}{\bibfnamefont{J.}~\bibnamefont{Rovny}},
  \bibinfo{author}{\bibfnamefont{R.~L.} \bibnamefont{Blum}}, \bibnamefont{and}
  \bibinfo{author}{\bibfnamefont{S.~E.} \bibnamefont{Barrett}},
  \bibinfo{journal}{Physical review letters} \textbf{\bibinfo{volume}{120}},
  \bibinfo{pages}{180603} (\bibinfo{year}{2018}{\natexlab{b}}).

\bibitem[{\citenamefont{Else et~al.}(2020)\citenamefont{Else, Monroe, Nayak,
  and Yao}}]{else2020discrete}
\bibinfo{author}{\bibfnamefont{D.~V.} \bibnamefont{Else}},
  \bibinfo{author}{\bibfnamefont{C.}~\bibnamefont{Monroe}},
  \bibinfo{author}{\bibfnamefont{C.}~\bibnamefont{Nayak}}, \bibnamefont{and}
  \bibinfo{author}{\bibfnamefont{N.~Y.} \bibnamefont{Yao}},
  \bibinfo{journal}{Annual Review of Condensed Matter Physics}
  \textbf{\bibinfo{volume}{11}}, \bibinfo{pages}{467} (\bibinfo{year}{2020}),
  \eprint{https://doi.org/10.1146/annurev-conmatphys-031119-050658},
  \urlprefix\url{https://doi.org/10.1146/annurev-conmatphys-031119-050658}.

\bibitem[{\citenamefont{Luitz et~al.}(2020)\citenamefont{Luitz, Moessner,
  Sondhi, and Khemani}}]{luitz2020prethermalization}
\bibinfo{author}{\bibfnamefont{D.~J.} \bibnamefont{Luitz}},
  \bibinfo{author}{\bibfnamefont{R.}~\bibnamefont{Moessner}},
  \bibinfo{author}{\bibfnamefont{S.}~\bibnamefont{Sondhi}}, \bibnamefont{and}
  \bibinfo{author}{\bibfnamefont{V.}~\bibnamefont{Khemani}},
  \bibinfo{journal}{Physical Review X} \textbf{\bibinfo{volume}{10}},
  \bibinfo{pages}{021046} (\bibinfo{year}{2020}).

\bibitem[{\citenamefont{Peng}(2019)}]{peng2019prethermalization}
\bibinfo{author}{\bibfnamefont{P.}~\bibnamefont{Peng}}, Ph.D. thesis,
  \bibinfo{school}{Massachusetts Institute of Technology}
  (\bibinfo{year}{2019}).

\bibitem[{\citenamefont{Peng et~al.}(2022{\natexlab{a}})\citenamefont{Peng, Ye,
  Yao, and Cappellaro}}]{peng2022disorder}
\bibinfo{author}{\bibfnamefont{P.}~\bibnamefont{Peng}},
  \bibinfo{author}{\bibfnamefont{B.}~\bibnamefont{Ye}},
  \bibinfo{author}{\bibfnamefont{N.~Y.} \bibnamefont{Yao}}, \bibnamefont{and}
  \bibinfo{author}{\bibfnamefont{P.}~\bibnamefont{Cappellaro}},
  \emph{\bibinfo{title}{Exploiting disorder to probe spin and energy
  hydrodynamics}} (\bibinfo{year}{2022}{\natexlab{a}}),
  \urlprefix\url{https://arxiv.org/abs/2209.09322}.

\bibitem[{\citenamefont{Martin et~al.}(2022)\citenamefont{Martin, Zhou, Leitao,
  Maskara, Makarova, Gao, Zhu, Park, Tyler, Park
  et~al.}}]{martin2022localtherm}
\bibinfo{author}{\bibfnamefont{L.~S.} \bibnamefont{Martin}},
  \bibinfo{author}{\bibfnamefont{H.}~\bibnamefont{Zhou}},
  \bibinfo{author}{\bibfnamefont{N.~T.} \bibnamefont{Leitao}},
  \bibinfo{author}{\bibfnamefont{N.}~\bibnamefont{Maskara}},
  \bibinfo{author}{\bibfnamefont{O.}~\bibnamefont{Makarova}},
  \bibinfo{author}{\bibfnamefont{H.}~\bibnamefont{Gao}},
  \bibinfo{author}{\bibfnamefont{Q.-Z.} \bibnamefont{Zhu}},
  \bibinfo{author}{\bibfnamefont{M.}~\bibnamefont{Park}},
  \bibinfo{author}{\bibfnamefont{M.}~\bibnamefont{Tyler}},
  \bibinfo{author}{\bibfnamefont{H.}~\bibnamefont{Park}}, \bibnamefont{et~al.},
  \emph{\bibinfo{title}{Controlling local thermalization dynamics in a
  floquet-engineered dipolar ensemble}} (\bibinfo{year}{2022}),
  \urlprefix\url{https://arxiv.org/abs/2209.09297}.

\bibitem[{\citenamefont{Leroy et~al.}(2001)\citenamefont{Leroy, Bres, Jones,
  and Downes}}]{leroy2001structure}
\bibinfo{author}{\bibfnamefont{N.}~\bibnamefont{Leroy}},
  \bibinfo{author}{\bibfnamefont{E.}~\bibnamefont{Bres}},
  \bibinfo{author}{\bibfnamefont{D.}~\bibnamefont{Jones}}, \bibnamefont{and}
  \bibinfo{author}{\bibfnamefont{S.}~\bibnamefont{Downes}},
  \bibinfo{journal}{European Cells and Materials} \textbf{\bibinfo{volume}{2}},
  \bibinfo{pages}{36} (\bibinfo{year}{2001}).

\bibitem[{\citenamefont{Zhang et~al.}(2009)\citenamefont{Zhang, Cappellaro,
  Antler, Pepper, Cory, Dobrovitski, Ramanathan, and Viola}}]{zhang2009nmr}
\bibinfo{author}{\bibfnamefont{W.}~\bibnamefont{Zhang}},
  \bibinfo{author}{\bibfnamefont{P.}~\bibnamefont{Cappellaro}},
  \bibinfo{author}{\bibfnamefont{N.}~\bibnamefont{Antler}},
  \bibinfo{author}{\bibfnamefont{B.}~\bibnamefont{Pepper}},
  \bibinfo{author}{\bibfnamefont{D.~G.} \bibnamefont{Cory}},
  \bibinfo{author}{\bibfnamefont{V.~V.} \bibnamefont{Dobrovitski}},
  \bibinfo{author}{\bibfnamefont{C.}~\bibnamefont{Ramanathan}},
  \bibnamefont{and} \bibinfo{author}{\bibfnamefont{L.}~\bibnamefont{Viola}},
  \bibinfo{journal}{Physical Review A} \textbf{\bibinfo{volume}{80}},
  \bibinfo{pages}{052323} (\bibinfo{year}{2009}).

\bibitem[{\citenamefont{Knill et~al.}(2000)\citenamefont{Knill, Laflamme,
  Martinez, and Tseng}}]{knill2000algorithmic}
\bibinfo{author}{\bibfnamefont{E.}~\bibnamefont{Knill}},
  \bibinfo{author}{\bibfnamefont{R.}~\bibnamefont{Laflamme}},
  \bibinfo{author}{\bibfnamefont{R.}~\bibnamefont{Martinez}}, \bibnamefont{and}
  \bibinfo{author}{\bibfnamefont{C.-H.} \bibnamefont{Tseng}},
  \bibinfo{journal}{Nature} \textbf{\bibinfo{volume}{404}},
  \bibinfo{pages}{368} (\bibinfo{year}{2000}).

\bibitem[{\citenamefont{McKay et~al.}(2017)\citenamefont{McKay, Wood, Sheldon,
  Chow, and Gambetta}}]{mckay2017efficient}
\bibinfo{author}{\bibfnamefont{D.~C.} \bibnamefont{McKay}},
  \bibinfo{author}{\bibfnamefont{C.~J.} \bibnamefont{Wood}},
  \bibinfo{author}{\bibfnamefont{S.}~\bibnamefont{Sheldon}},
  \bibinfo{author}{\bibfnamefont{J.~M.} \bibnamefont{Chow}}, \bibnamefont{and}
  \bibinfo{author}{\bibfnamefont{J.~M.} \bibnamefont{Gambetta}},
  \bibinfo{journal}{Physical Review A} \textbf{\bibinfo{volume}{96}},
  \bibinfo{pages}{022330} (\bibinfo{year}{2017}).

\bibitem[{\citenamefont{Wei et~al.}(2018)\citenamefont{Wei, Ramanathan, and
  Cappellaro}}]{wei2018exploring}
\bibinfo{author}{\bibfnamefont{K.~X.} \bibnamefont{Wei}},
  \bibinfo{author}{\bibfnamefont{C.}~\bibnamefont{Ramanathan}},
  \bibnamefont{and}
  \bibinfo{author}{\bibfnamefont{P.}~\bibnamefont{Cappellaro}},
  \bibinfo{journal}{Physical review letters} \textbf{\bibinfo{volume}{120}},
  \bibinfo{pages}{070501} (\bibinfo{year}{2018}).

\bibitem[{\citenamefont{S\'anchez et~al.}(2020)\citenamefont{S\'anchez,
  Chattah, Wei, Buljubasich, Cappellaro, and Pastawski}}]{Sanchez20}
\bibinfo{author}{\bibfnamefont{C.~M.} \bibnamefont{S\'anchez}},
  \bibinfo{author}{\bibfnamefont{A.~K.} \bibnamefont{Chattah}},
  \bibinfo{author}{\bibfnamefont{K.~X.} \bibnamefont{Wei}},
  \bibinfo{author}{\bibfnamefont{L.}~\bibnamefont{Buljubasich}},
  \bibinfo{author}{\bibfnamefont{P.}~\bibnamefont{Cappellaro}},
  \bibnamefont{and} \bibinfo{author}{\bibfnamefont{H.~M.}
  \bibnamefont{Pastawski}}, \bibinfo{journal}{Phys. Rev. Lett.}
  \textbf{\bibinfo{volume}{124}}, \bibinfo{pages}{030601}
  (\bibinfo{year}{2020}),
  \urlprefix\url{https://link.aps.org/doi/10.1103/PhysRevLett.124.030601}.

\bibitem[{\citenamefont{Peng et~al.}(2022{\natexlab{b}})\citenamefont{Peng,
  Huang, Yin, Joseph, Ramanathan, Cappellaro et~al.}}]{peng2022deep}
\bibinfo{author}{\bibfnamefont{P.}~\bibnamefont{Peng}},
  \bibinfo{author}{\bibfnamefont{X.}~\bibnamefont{Huang}},
  \bibinfo{author}{\bibfnamefont{C.}~\bibnamefont{Yin}},
  \bibinfo{author}{\bibfnamefont{L.}~\bibnamefont{Joseph}},
  \bibinfo{author}{\bibfnamefont{C.}~\bibnamefont{Ramanathan}},
  \bibinfo{author}{\bibfnamefont{P.}~\bibnamefont{Cappellaro}},
  \bibnamefont{et~al.}, \bibinfo{journal}{Physical Review Applied}
  \textbf{\bibinfo{volume}{18}}, \bibinfo{pages}{024033}
  (\bibinfo{year}{2022}{\natexlab{b}}).

\bibitem[{\citenamefont{Zeng and Sheng}(2017)}]{zeng2017prethermal}
\bibinfo{author}{\bibfnamefont{T.-S.} \bibnamefont{Zeng}} \bibnamefont{and}
  \bibinfo{author}{\bibfnamefont{D.}~\bibnamefont{Sheng}},
  \bibinfo{journal}{Physical Review B} \textbf{\bibinfo{volume}{96}},
  \bibinfo{pages}{094202} (\bibinfo{year}{2017}).

\bibitem[{\citenamefont{Pizzi et~al.}(2021)\citenamefont{Pizzi, Knolle, and
  Nunnenkamp}}]{pizzi2021higher}
\bibinfo{author}{\bibfnamefont{A.}~\bibnamefont{Pizzi}},
  \bibinfo{author}{\bibfnamefont{J.}~\bibnamefont{Knolle}}, \bibnamefont{and}
  \bibinfo{author}{\bibfnamefont{A.}~\bibnamefont{Nunnenkamp}},
  \bibinfo{journal}{Nature communications} \textbf{\bibinfo{volume}{12}},
  \bibinfo{pages}{1} (\bibinfo{year}{2021}).

\bibitem[{\citenamefont{Giachetti et~al.}(2022)\citenamefont{Giachetti,
  Solfanelli, Correale, and Defenu}}]{giachetti2022fractal}
\bibinfo{author}{\bibfnamefont{G.}~\bibnamefont{Giachetti}},
  \bibinfo{author}{\bibfnamefont{A.}~\bibnamefont{Solfanelli}},
  \bibinfo{author}{\bibfnamefont{L.}~\bibnamefont{Correale}}, \bibnamefont{and}
  \bibinfo{author}{\bibfnamefont{N.}~\bibnamefont{Defenu}},
  \emph{\bibinfo{title}{High-order time crystal phases and their fractal
  nature}} (\bibinfo{year}{2022}),
  \urlprefix\url{https://arxiv.org/abs/2203.16562}.

\bibitem[{\citenamefont{Boutis et~al.}(2004)\citenamefont{Boutis, Greenbaum,
  Cho, Cory, and Ramanathan}}]{boutis2004spin}
\bibinfo{author}{\bibfnamefont{G.}~\bibnamefont{Boutis}},
  \bibinfo{author}{\bibfnamefont{D.}~\bibnamefont{Greenbaum}},
  \bibinfo{author}{\bibfnamefont{H.}~\bibnamefont{Cho}},
  \bibinfo{author}{\bibfnamefont{D.}~\bibnamefont{Cory}}, \bibnamefont{and}
  \bibinfo{author}{\bibfnamefont{C.}~\bibnamefont{Ramanathan}},
  \bibinfo{journal}{Physical review letters} \textbf{\bibinfo{volume}{92}},
  \bibinfo{pages}{137201} (\bibinfo{year}{2004}).

\bibitem[{\citenamefont{Zhang and Cory}(1998)}]{zhang1998first}
\bibinfo{author}{\bibfnamefont{W.}~\bibnamefont{Zhang}} \bibnamefont{and}
  \bibinfo{author}{\bibfnamefont{D.}~\bibnamefont{Cory}},
  \bibinfo{journal}{Physical review letters} \textbf{\bibinfo{volume}{80}},
  \bibinfo{pages}{1324} (\bibinfo{year}{1998}).

\bibitem[{\citenamefont{Peng et~al.}(2019)\citenamefont{Peng, Li, Yan, Wei,
  Cappellaro et~al.}}]{peng2019comparing}
\bibinfo{author}{\bibfnamefont{P.}~\bibnamefont{Peng}},
  \bibinfo{author}{\bibfnamefont{Z.}~\bibnamefont{Li}},
  \bibinfo{author}{\bibfnamefont{H.}~\bibnamefont{Yan}},
  \bibinfo{author}{\bibfnamefont{K.~X.} \bibnamefont{Wei}},
  \bibinfo{author}{\bibfnamefont{P.}~\bibnamefont{Cappellaro}},
  \bibnamefont{et~al.}, \bibinfo{journal}{Physical Review B}
  \textbf{\bibinfo{volume}{100}}, \bibinfo{pages}{214203}
  (\bibinfo{year}{2019}).

\bibitem[{\citenamefont{Serbyn et~al.}(2013)\citenamefont{Serbyn, Papi{\'c},
  and Abanin}}]{serbyn2013local}
\bibinfo{author}{\bibfnamefont{M.}~\bibnamefont{Serbyn}},
  \bibinfo{author}{\bibfnamefont{Z.}~\bibnamefont{Papi{\'c}}},
  \bibnamefont{and} \bibinfo{author}{\bibfnamefont{D.~A.}
  \bibnamefont{Abanin}}, \bibinfo{journal}{Physical review letters}
  \textbf{\bibinfo{volume}{111}}, \bibinfo{pages}{127201}
  (\bibinfo{year}{2013}).

\bibitem[{\citenamefont{Serbyn et~al.}(2014)\citenamefont{Serbyn, Papi{\'c},
  and Abanin}}]{serbyn2014quantum}
\bibinfo{author}{\bibfnamefont{M.}~\bibnamefont{Serbyn}},
  \bibinfo{author}{\bibfnamefont{Z.}~\bibnamefont{Papi{\'c}}},
  \bibnamefont{and} \bibinfo{author}{\bibfnamefont{D.~A.}
  \bibnamefont{Abanin}}, \bibinfo{journal}{Physical Review B}
  \textbf{\bibinfo{volume}{90}}, \bibinfo{pages}{174302}
  (\bibinfo{year}{2014}).

\bibitem[{\citenamefont{Virtanen et~al.}(2020)\citenamefont{Virtanen, Gommers,
  Oliphant, Haberland, Reddy, Cournapeau, Burovski, Peterson, Weckesser, Bright
  et~al.}}]{2020SciPy-NMeth}
\bibinfo{author}{\bibfnamefont{P.}~\bibnamefont{Virtanen}},
  \bibinfo{author}{\bibfnamefont{R.}~\bibnamefont{Gommers}},
  \bibinfo{author}{\bibfnamefont{T.~E.} \bibnamefont{Oliphant}},
  \bibinfo{author}{\bibfnamefont{M.}~\bibnamefont{Haberland}},
  \bibinfo{author}{\bibfnamefont{T.}~\bibnamefont{Reddy}},
  \bibinfo{author}{\bibfnamefont{D.}~\bibnamefont{Cournapeau}},
  \bibinfo{author}{\bibfnamefont{E.}~\bibnamefont{Burovski}},
  \bibinfo{author}{\bibfnamefont{P.}~\bibnamefont{Peterson}},
  \bibinfo{author}{\bibfnamefont{W.}~\bibnamefont{Weckesser}},
  \bibinfo{author}{\bibfnamefont{J.}~\bibnamefont{Bright}},
  \bibnamefont{et~al.}, \bibinfo{journal}{Nature Methods}
  \textbf{\bibinfo{volume}{17}}, \bibinfo{pages}{261} (\bibinfo{year}{2020}).

\end{thebibliography}
\end{document}